\documentclass[structabstract]{aa}  
\usepackage{graphicx}
\begin{document}
   \title{Star formation in the intragroup medium and other diagnostics of the evolutionary stages of compact groups of galaxies\thanks{Based on observations collected at the European Southern Observatory, La Silla, Chile}}

   \subtitle{}

   \author{S. Torres-Flores
          \inst{1,2},
	  C. Mendes de Oliveira\inst{1}, 
          D. F. de Mello\inst{3,4,5}, 
          P. Amram\inst{2},
          H. Plana\inst{2,7}
	  B. Epinat\inst{6},
          \and
	  J. Iglesias-P\'aramo\inst{8}
          }

   \institute{Universidade de S\~ao Paulo, Instituto de Astronomia, 
              Geof\' isica e Ci\^encias Atmosf\'ericas, Departamento de
              Astronomia, S\~ao Paulo, Brazil
         \and
             Laboratoire d'Astrophysique de Marseille, OAMP, Universit\'e de Provence \& CNRS, 38 rue F. Joliot--Curie, 13388 Marseille, Cedex 13, France
         \and
              Observational Cosmology Laboratory, Code 665, Goddard Space Flight Center, Greenbelt, MD 20771, USA
         \and
              Catholic University of America, Washington, DC 20064, USA
         \and
              Johns Hopkins University, Baltimore, MD 21218, USA
         \and
              Laboratoire d'Astrophysique de Toulouse-Tarbes, Universit\'e de Toulouse,
              CNRS, 14 Avenue Edouard Belin, 31400 Toulouse, France 
	 \and
              Laboratorio de Astrofisica Teorica e Observacional, Universidade Estadual de Santa Cruz, Brazil
         \and
             Instituto de Astrofisica de Andalucia (CSIC), Camino Bajo de Huetor 50, 18008 Granada, Spain
             }

   \date{Received      ; accepted      }

% \abstract{}{}{}{}{} 
% 5 {} token are mandatory
 
  \abstract
   {Compact groups of galaxies are entities that have high densities of galaxies and serve as laboratories to study galaxy interactions, intergalactic star formation and galaxy evolution.} 
%This phenomena could trigger star formation in the intragroup medium, in the form of intergalactic HII regions or tidal dwarf galaxies.}
   {The main goal of this study is to search for 
young objects in the intragroup medium of seven compact groups of galaxies: HCG 2, 7, 22, 23, 92, 100 and NGC 92 as well as to evaluate the stage of interaction of each group.}
   {We used Fabry-Perot velocity fields and rotation curves 
together with \textit{GALEX} NUV and FUV images and optical R-band and HI maps.}
   {(i) HCG 7 and HCG 23 are in early stages of interaction, (ii) HCG 2 and HCG 22 are mildly interacting, and (iii) HCG 92, HCG 100 and NGC 92 are in late 
stages of evolution.
We find that all three evolved groups contain populations of young blue objects in the intragroup medium, consistent 
with ages $<$ 100 Myr, of which several are younger than $<$ 10 Myr. We also report the discovery of a tidal dwarf galaxy candidate in the tail of NGC 92. These three groups, besides containing galaxies that have peculiar velocity fields, also 
show extended HI tails.}
   {Our results indicate that the advanced stage of evolution of a group, together with the presence of intragroup HI clouds, may lead to star formation in the intragroup medium. A table containing all intergalactic HII regions and tidal dwarf galaxies confirmed to date is appended.}

\titlerunning{Dynamic stages of compact groups of galaxies}
\authorrunning{S. Torres-Flores et al.}

   \maketitle
%
%________________________________________________________________

\section{Introduction}

It is well known that tidal interactions can alter galaxies' properties and, in particular, the neutral gas can be plucked by tidal forces and develop tidal tails where young stellar clusters, HII regions, and tidal dwarf galaxies (TDGs)
may form (e.g. de Mello et al. 2008b, de Mello, Torres-Flores, \& Mendes de Oliveira 2008a, 
Mendes de Oliveira et al. 2004, Ryan-Weber et al. 2004, Mendes de Oliveira et al. 2001 and Duc \& Mirabel 1998).
However, the fate of these young systems remain unclear (de Mello et al. 2008b) and they can  
either (1) become independent entities growing by accreting 
more gas and forming more stars to be transformed in TDGs, (2) make star clusters that survive and migrate in the distant halos 
of their hosts or even become intergalactic objects, or (3) dissolve and not remain gravitationally 
bound, yielding only very sparse star streams.

Hunsberger et al. (1996) suggested  that at least 
one-third and perhaps more than one-half of all dwarf galaxies in compact groups are formed 
during galaxy interactions. Therefore, a considerable number of new stellar systems 
may have formed and may be linked to the evolutionary stage of the group. There are several 
diagnostics used in the literature to evaluate these stages. 
For instance, Coziol et al. (2004) developed a scheme 
to classify the evolutionary stage of compact groups of galaxies based on the morphologies of the members and their level of activity. 
Plana et al. (2003) and Amram et al. (2003), using 
2-D H$\alpha$ velocity fields, classified several of the Hickson compact groups in 
three classes: unevolved group, mildly interacting group, 
and a system in the final stage of evolution. 
Verdes-Montenegro et al. (2001), on the other hand, used the HI content of compact groups to 
propose an evolutionary sequence. In their classification, groups may show HI still attached to the 
individual members, in tidal tails or a in common 
HI envelope around all galaxies.
In this context, HI and kinematic information from 2D-velocity maps have proved to be the most useful 
tools to study the evolutionary stages of groups. 

In this paper we present the results of a multiwavelength campaign we have conducted using
several instruments in order to properly evaluate the evolutionary stages of seven compact groups 
of galaxies. We report the discovery of young objects in the intragroup medium of compact groups and in the tidal tails of strongly interacting galaxies, which may have formed due to galaxy
interactions, using
\textit{GALEX}/UV data, optical R-images, new Fabry 
Perot velocity maps and HI maps from the literature. UV images reveal important information regarding the
age of the young objects while the velocity 
fields and rotation curves of the group members help constraining the 
evolutionary stage of each compact group.

This paper is organized as follows: In \S 2 we present the data. In \S 3 we present the data analysis. In \S 4 we present the ultraviolet and Fabry Perot results. In \S 5 we discuss our results and 
in \S 6 we present our main conclusions.

%__________________________________________________________________

\section{Data}

%                                     Two column figure (place early!)
%______________________________________________ Gamma_1 (lg rho, lg e)

%__________________________________________________ One column table

%
%                                                One column figure
%----------------------------------------------------------- S_vib
  
%
%______________________________________________________________

In the following sections we describe the data we have obtained 
for a sample of seven compact groups of galaxies: 
\object{HCG 2}, \object{HCG 7}, \object{HCG 22}, \object{HCG 23}, 
\object{HCG 92}, \object{HCG 100} and \object{NGC 92}. These nearby groups (with velocities lower than 7000 km/s) span a wide
range of properties (density, isolation, spiral fraction and HI content. See Hickson 1982 and Verdes Montenegro et al. 2001).
In particular isolated groups may evolve differently from groups which are
parts of larger structures (such as HCG 23, Williams \& van Gorkom 1995). 
On the other hand, HCG 92
and HCG 100 are known to have HI emission in the intragroup medium (Verdes-Montenegro et al. 2001 and de Mello et al. 2008a, respectively) while
NGC 92 has a common halo of HI around the whole group (Pompei et al. 2007) and the others have
HI emission which is more concentrated on the galaxies.  Therefore, the
groups chosen are representative of the general compact group population.
In Table \ref{table1}, we summarize the main properties of this sample.

\subsection{Fabry-Perot Data}

Observations of HCG 2, HCG 7, HCG 22 and NGC 92  
were carried out at the European Southern Observatory 3.6 m telescope (ESO) in September 2000. 
The data were obtained with the Fabry Perot instrument CIGALE using an Image Photon Counting System, for a 
field of view of 4 arcmin centered on each group. These data are
described in detail in Torres-Flores et al. (2009). 
The scale of the images was 0.405\arcsec/pixel. The data were
taken under non-photometric conditions and seeing of 1\arcsec. No
flux calibration was attempted. Data reduction was performed using
a software developed by Daigle et al. (2006, see also Epinat et
al. 2008a). The data for HCG 100 was taken from Plana et
al. (2003).

The output of the reduction package
gave us 2D monochromatic, continuum and radial velocity maps, with a signal-to-noise higher than 6. Wavelength calibration was obtained by scanning the 
narrow Ne 6599 \AA{} line under the same conditions as the observations. 
The velocity sampling was 12 km s$^{-1}$. 
For HCG 2, 7 and 22, the OH subtraction was performed by estimating a sky cube from the sky dominated regions and subtracting it from the original cube. In the case of NGC 92, the OH was subtracted estimating the sky using the medium spectrum of the data cube. This method resulted in a better velocity field for NGC 92 than when the sky cube method was adopted.

\subsection{Ultraviolet Data}

The UV images of the seven compact groups in this sample were obtained using the \textit{Galaxy Evolution Explorer} 
(\textit{GALEX}) satellite in the near ultraviolet (NUV $\lambda$$_{\rm eff}$=2271\AA) and 
far ultraviolet (FUV $\lambda$$_{\rm eff}$=1528\AA) bands (Cycle 1, program number 73 and 31 and 
public archival data). 
The images analyzed herein are the intensity maps with the background removed (given by the 
\textit{GALEX} pipeline). FUV and NUV fluxes were calculated using Morrissey et al. (2005) 
m$_{\lambda}$=-2.5 log[F$_{\lambda}$/a$_{\lambda}$] + b$_{\lambda}$, 
where a$_{FUV}$ = 1.4 $\times$ 10$^{-15}$ erg s$^{-1}$ cm$^{-2}$ \AA$^{-1}$,  
a$_{NUV}$=2.06$\times$ 10$^{-16}$ erg s$^{-1}$ cm$^{-2}$ \AA$^{-1}$, b$_{FUV}$=18.82 and b$_{NUV}$=20.08 for FUV and NUV, respectively. Fluxes were multiplied
by the effective filter bandpass ($\Delta$$\lambda$$_{FUV}$=269\AA\, and $\Delta$$\lambda$$_{NUV}$=616\AA) to give 
units of erg s$^{-1}$ cm$^{-2}$. The \textit{GALEX} fields view are 1$^{\circ}$.28 and 1$^{\circ}$.24 in FUV and NUV respectively and the pixel scale is 1.5 arcsec pixel$^{-1}$.

HCG 100 and HCG 92 were originally analyzed in de Mello et al. (2008a)\footnote{de Mello et al. (2008a) selected the UV sources based on proximity to the HI tail while in this work, sources were 
selected within a region of radius equal to two times the radius of the minimum circle, as defined in Session 3.2} and Xu et al. (2005), respectively, and 
they have been re-analyzed here using the same method applied to the other groups. Tzanavaris et al. (2009) used SWIFT/UVOT to analyze the UV properties of another sample of Hickson Compact Groups of galaxies which will be complementary to our GALEX data. 

\subsection{R-band Data}
R-band images of HCG 2, 23 and 100 were obtained at the Cerro Tololo Interamerican Observatory on Aug 23/2005,
using the 4 meter Blanco telescope and a mosaic II CCD imager, with a pixel scale of 0.27 arcsec/pixel, for 
a field of view of 40 $\times$ 40 arcmin, centered on each group. For each target, three exposures of
300s were taken in the R band, with a typical seeing of 1.1 arcsec. The data were bias subtracted
and flat-fielded using IRAF\footnote{IRAF is distributed by the National Optical Astronomy Observatories, which are 
operated by the Association of Universities for Research in Astronomy, Inc., under cooperative 
agreement with the National Science Foundation.}, using standard procedures (package mscred). A flat-field was constructed 
from a combination of dark sky frames and twilight flats which worked well, given that the background,
after flat-fielding, showed a rms variation lower than 1\% over the whole field.
Since the night was not photometric, no calibration stars were observed and the zeropoints were obtained from 
published surface brightness profiles of group members 
(Mendes de Oliveira 1992, Rubin et al. 1991).

HCG 22 was observed with the Keck II telescope in the R and B bands. These data are described in da Rocha et al. (2002).

R-band image for NGC 92 was obtained at La Silla Observatory with the NTT telescope. It was taken on Oct 27th 1994, within a program to monitor 
the supernova SN1994Z, which was discovered in October 1994, 
in the galaxy NGC 87, a close companion to NGC 92. 
The seeing was 1$''$ and the pixel scale was 0.27 $''$/pixel. The zeropoint was estimated with calibration stars taken in the same night as the targets. 

HCG 92 was observed with Gemini in the Sloan filters g, r and i. These data are described in Mendes de Oliveira et al. (2004).
For HCG 7, the r-band image of the
Sloan Digital Sky Survey (SDSS) was used in our analysis. Transformation between 
r-band magnitudes and R-band magnitudes was performed using R = r - 0.1837$\times$(g - r) (Lupton 2005) for the sources in groups HCG 7 and 92.

\begin{table*}
\begin{minipage}[t]{\textwidth}
\caption{Properties of the Sample}
\label{table1}
\centering
\renewcommand{\footnoterule}{}  % to avoid a line before footnotes
\begin{tabular}{lcccccccccc}
\hline \hline
    ~     & \multicolumn{5}{c}{} & \multicolumn{3}{c}{POSITION ANGLE (PA (deg))} & \multicolumn{2}{c}{INCLINATION (deg)} \\
Name & Morph.\footnote{Taken from Hickson (1993) and from NED database for NGC 92. Morphological type for NGC 92 was taken from Prugniel et al. (1998).} & Distance\footnote{Distances were obtained using the redshift for each group given in Hickson (1992) and assuming the groups follow the Hubble flow. For NGC 92, we used the mean redshift of the group members. For each galaxy, redshift was taken from NED. In all cases, we used H$_{0}$=70 km s$^{-1}$ Mpc$^{-1}$, $\Omega_{m}$=0.3 and $\Omega_{\lambda}$=0.7.} & B$_{TC}$\footnotesize$^{a}$ & V$_{sys}$\footnotesize$^{a}$ &V$_{sys}$\footnote{Systemic velocity deduced from our velocity field analysis.} & Velocity & Optical\footnote{For HCG 2a, b, c, HCG 7a, d, HCG 22c, and values of PA were taken from Mendes de Oliveira (1992). These were determined 
at semimajor axis lengths of 2\textit{h}$^{-1}$kpc for all Hickson group galaxies with z$\leq$0.05. For NGC 92 the PA was 
taken from the Hyperleda database.} & Optical\footnote{Taken from the Hyperleda database. PA is the position angle of the 
major axis at the 25 mag arcsec$^{-2}$ isophote, in the B-band} &Velocity & Optical \\ ~ & Type & Mpc & mag & km s$^{-1}$ & km s$^{-1}$ & Map & image & image & Map & image \\
\hline
HCG 2\\
a   &  SBd & 62.36 & 13.35  &   4326 & 4347  & 41$\pm$3  & 3 & 8 & 68$\pm$13 & 61 \\
b   &  cI  & 62.36 & 14.39  &   4366 & 4357  & 8$\pm$3  & 28 & 3 & 49$\pm$12 & 30 \\
c   &  SBc & 62.36 & 14.15  &   4235 & 4266  & 168$\pm$3 & 132 & 158 & ... & 57\\
HCG 7\\
a   & Sb   & 61.04 & 12.98  & 4210 & 4198  & 176$\pm$1  &  153   &  165  & 67$\pm$3 & 61\\
b   & SB0  & 61.04 & 13.74  & 4238 &   &  &	&    &  	& \\
c\footnote{Taken from Mendes de Oliveira et al. (2003).}   & SBc  & 61.04 & 12.60  &  4366 & &  133  & 133 & 144 & 48 & 36 \\
d   & SBc  & 61.04 & 14.77  & 4116 & 4104  & 85$\pm$3  &  58   &  35  &  31$\pm$12 & 43 \\
HCG 22\\
a  &   E2  & 38.81 &  12.24  & 2705 & &    &    &   &    & \\
b  &   Sa  & 38.81 &  14.47  & 2625 & &    &    &   &    & \\
c  & SBcd  & 38.81 & 13.90   & 2728 & 2572  & 101$\pm$2  &  70  &  179 & ... & 25\\
HCG 23\\
a  &  Sab & 69.81 & 14.32  & 4798  & & &  &  &  & \\
b  &  SBc & 69.81 & 14.42  & 4921  & & &  &  &  & \\
c  &  S0  & 69.81 & 15.52  & 5016  & & &  &  &  & \\
d  &  Sd  & 69.81 & 16.00  & 4562  & & &  &  &  & \\
HCG 92\\
b & Sbc  & 93.60 & 13.18 & 5774 & & &  &  &  & \\
c & SBc  & 93.60 & 13.33 & 6764 & & &  &  &  & \\
d & Sc   & 93.60 & 13.63 & 6630 & & &  &  &  & \\
e & E1   & 93.60 & 14.01 & 6599 & & &  &  &  & \\
HCG 100\footnote{Taken from Plana et al. (2003).}\\
a &  Sb   & 77.28 & 13.66 & 5300&  5323 & 78$\pm$5  & 78  & 79 &50$\pm$7 &47\\
b &  Sm   & 77.28 & 14.90 & 5253&  5163 & 145$\pm$5  & 165 & 149 &52$\pm$10 &57\\
c &  SBc  & 77.28 & 15.22 & 5461&  5418 & 72$\pm$8 & 70  & 67 &66$\pm$5 &65\\
d &  Scd  & ...   & 15.97 & ... &   & 53$\pm$3  & 47  & 50 &70$\pm$5 &64\\
NGC 92\\
NGC 92  & Sa        & 48.50 & ... & 3219 & 3412 &  148$\pm$3 &  149   & 149   &48$\pm$6 &62\\
NGC 89  & SB(s)0/a   & 48.50 & ... & 3320 & &   &  &  & & \\
NGC 88  & SB(rs)0/a  & 48.50 & ... & 3433 & &   &  &  & & \\
NGC 87  & IBm        & 48.50 & ... & 3491 & &   &  &  & & \\
\hline
\end{tabular}
\end{minipage}
\end{table*}

\section{Data Analysis}

\subsection{Rotation Curves and Kinematic Parameters}

In order to constrain the evolutionary stage of each compact group, we inspected the velocity field and rotation curve of each 
member to 
search for interaction 
indicators\footnote{Early-type galaxies HCG 7b and HCG 22b (Table \ref{table1}) had very weak H$\alpha$ emission  
and, therefore, no rotation curves were derived for these objects.}.
Rotation curves were constructed using the morphological center 
(defined as the location of the peak intensity in the continuum image) instead of the kinematic center 
(defined as the center that produces a symmetric rotation curve) because in most cases the 
rotation curves were asymmetric. Asymmetry made it difficult to match both sides of the curve, no matter which center was chosen. 
In most cases there was clearly a decoupling between the morphological and kinematic centers (see Amram et al. 2007). 
Kinematic position angles, inclinations, systemic velocities (see Table \ref{table1}) and rotation curves were derived from the velocity fields using the procedure developed by Epinat et al. (2008a). The position of the morphological 
center for each galaxy is marked with a plus sign in the velocity fields
of Figs. 1 to 6.

The optical position angles (PAs) were taken from Mendes de Oliveira (1992) and from the Hyperleda\footnote{http://leda.univ-lyon1.fr} database.

In all cases, the PA is oriented from North to East. 
For HCG 2c and HCG 22c, the kinematic inclination resulted in unrealistic values which resulted in maximum rotation velocities higher than the expected velocities (Torres-Flores et al. 2009). Therefore, 
in these cases, we used the morphological inclinations obtained from the relation cos(\textit{b}/\textit{a})=i, where \textit{b} and \textit{a} correspond 
to the length of the minor and major axis of the $\mu$$_{B}$=25 mag arc$^{-2}$ isophote given in Hickson (1993). 

\begin{figure*}[t!]
\includegraphics[width=\textwidth]{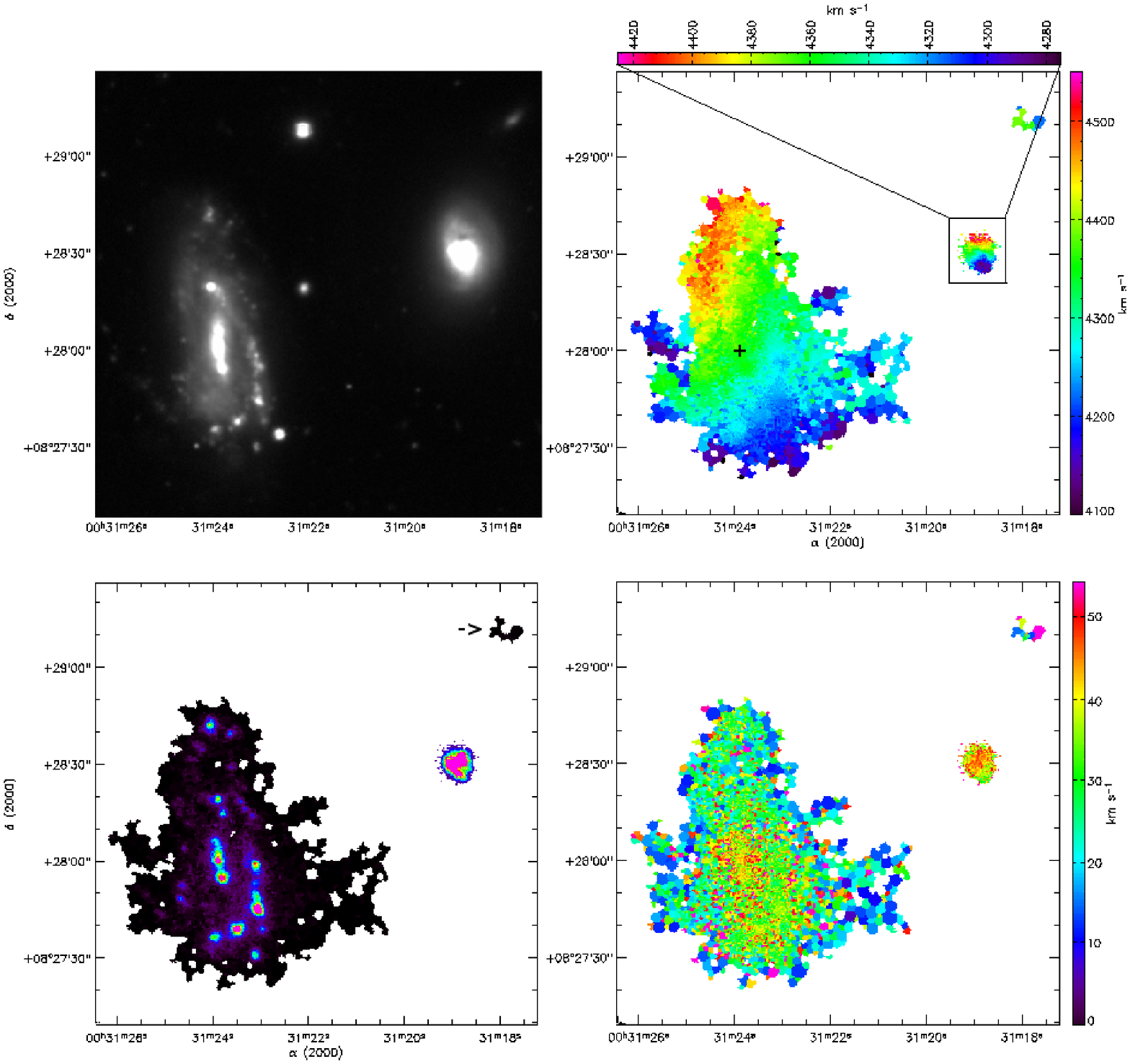}
\caption{HCG 2a,b. North is at the top and East is to the left of all images in Figures \ref{maps_hcg2ab} to \ref{maps_s01a}. HCG 2a is located to the East of HCG 2b. 
Top left: R band image from CTIO. Top right: Velocity field. We highlight HCG 2b own scale at the top of the velocity field, 
in order to show clearly its dynamic range. Bottom left: Monochromatic image. 
Bottom right: Velocity dispersion map. Small arrow in the R band image and monochromatic map indicates the position of a new source (probably 
in the background) detected in the Fabry-Perot analysis.}
\label{maps_hcg2ab}
\end{figure*}
%\clearpage

\begin{figure*}[t!]
\includegraphics[width=\textwidth]{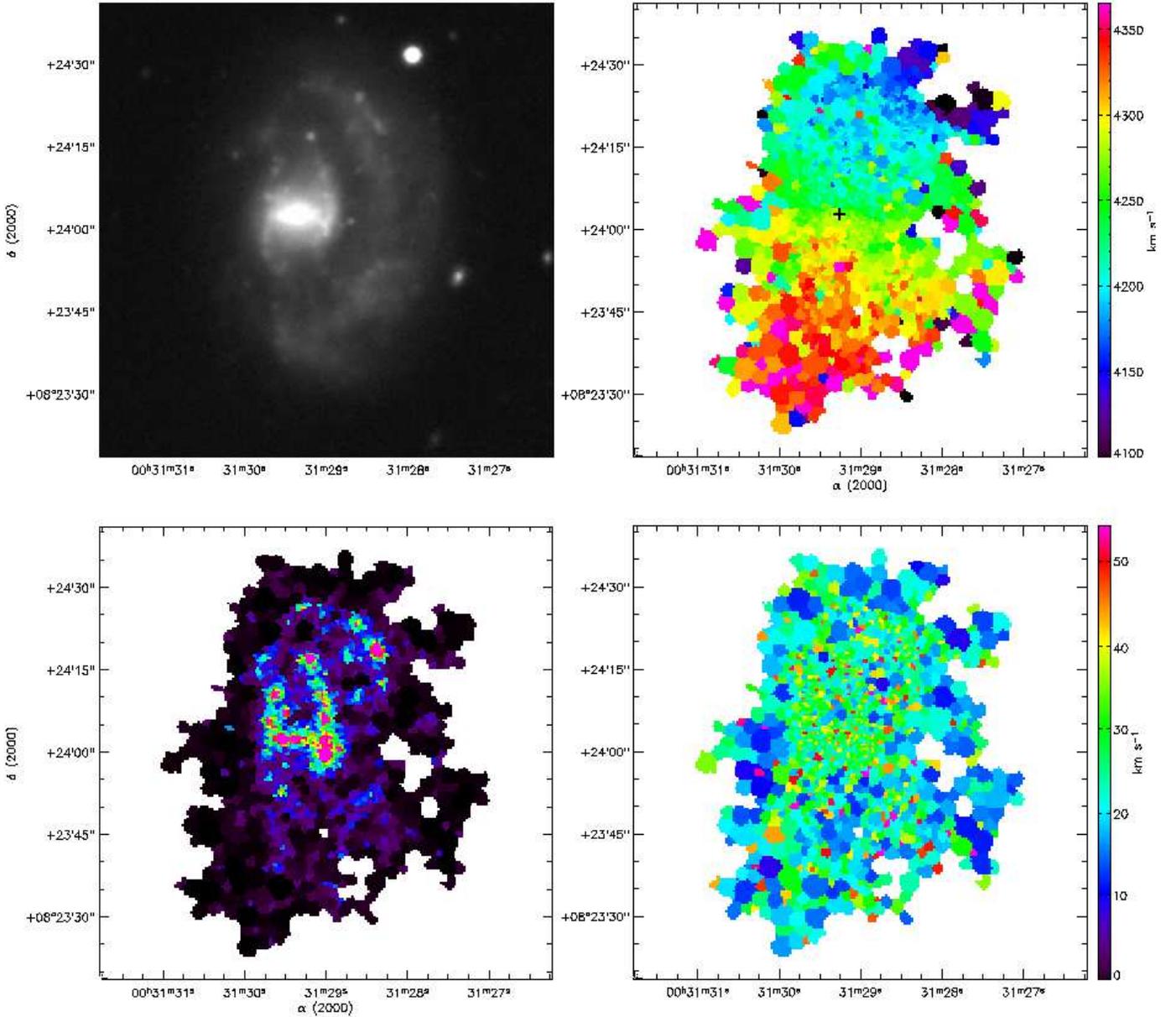}
\caption{HCG 2c. Top left: R band image from CTIO. Top right: Velocity field. Bottom left: Monochromatic image. Bottom right: Velocity dispersion map.}
\label{maps_hcg2c}
\end{figure*}
%\clearpage

\begin{figure*}[t!]
\includegraphics[width=\textwidth]{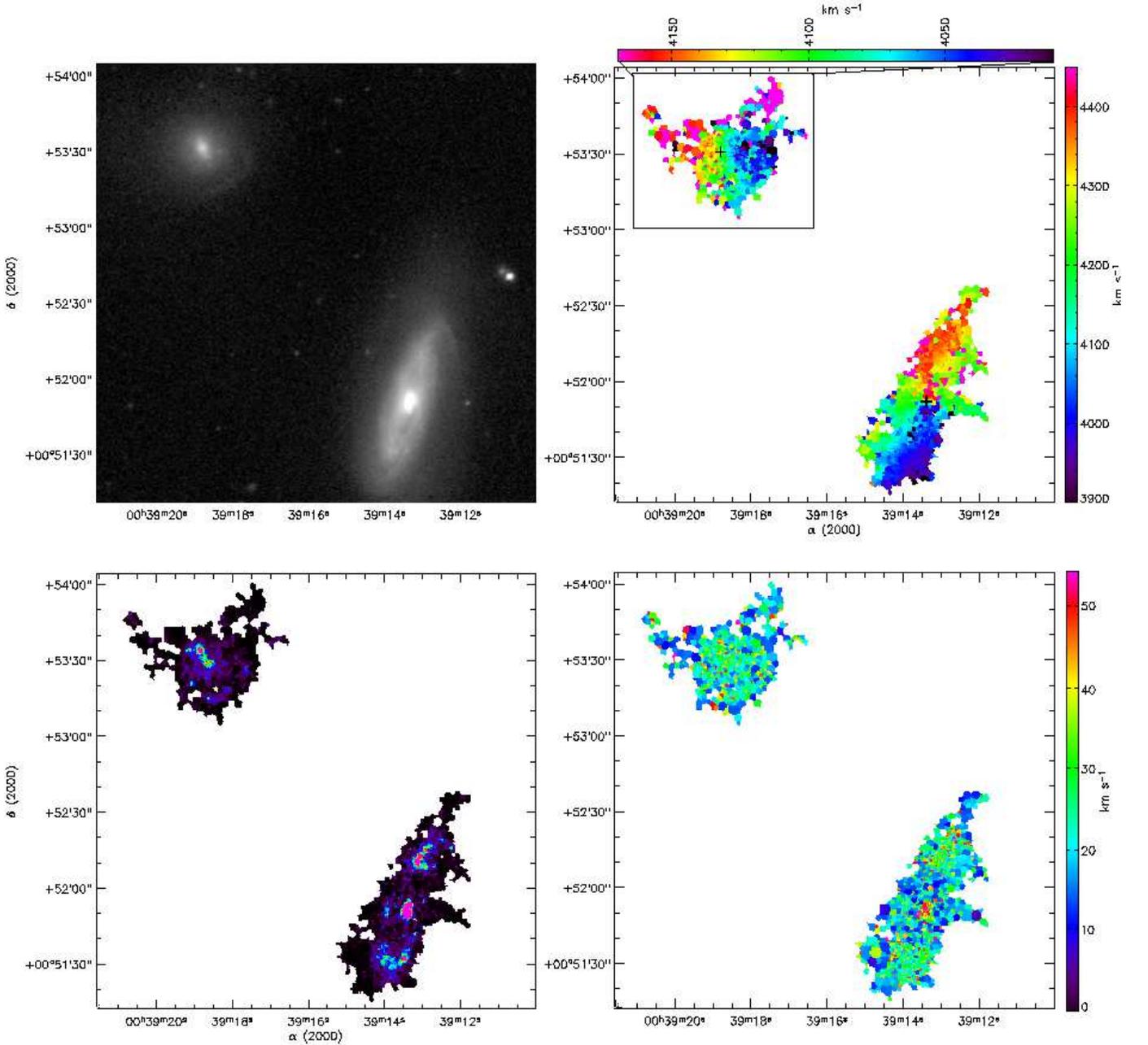}
\caption{HCG 7a,d. HCG 7a is located Southwest of HCG 7d. Top left: r band image from SDSS. Top right: Velocity field. 
We highlight HCG 7d own scale at the top of the velocity field, 
in order to show clearly its dynamic range. 
Bottom left: Monochromatic image. Bottom right: Velocity dispersion map.}
\label{maps_hcg7ad}
\end{figure*}
%\clearpage

\begin{figure*}[t!]
\includegraphics[width=\textwidth]{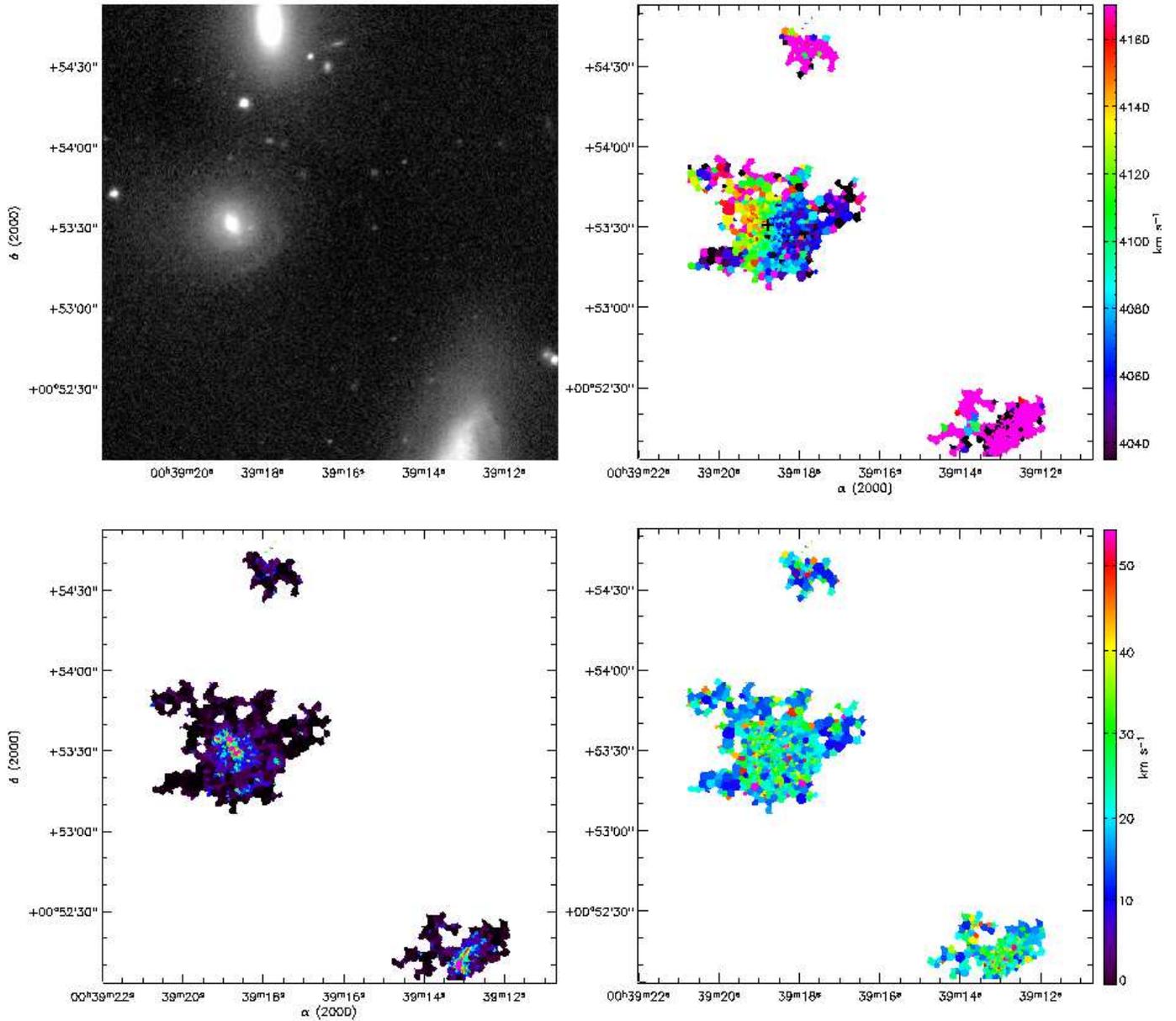}
\caption{HCG 7b,d. HCG 7d is located in the center of the field. HCG 7a and b are partially shown in the South and North of the field respectively. 
Top left: r band image from SDSS. Top right: Velocity field. Bottom left: Monochromatic image. Bottom right: Velocity dispersion map.}
\label{maps_hcg7bd}
\end{figure*}
%\clearpage

\begin{figure*}[t!]
\includegraphics[width=\textwidth]{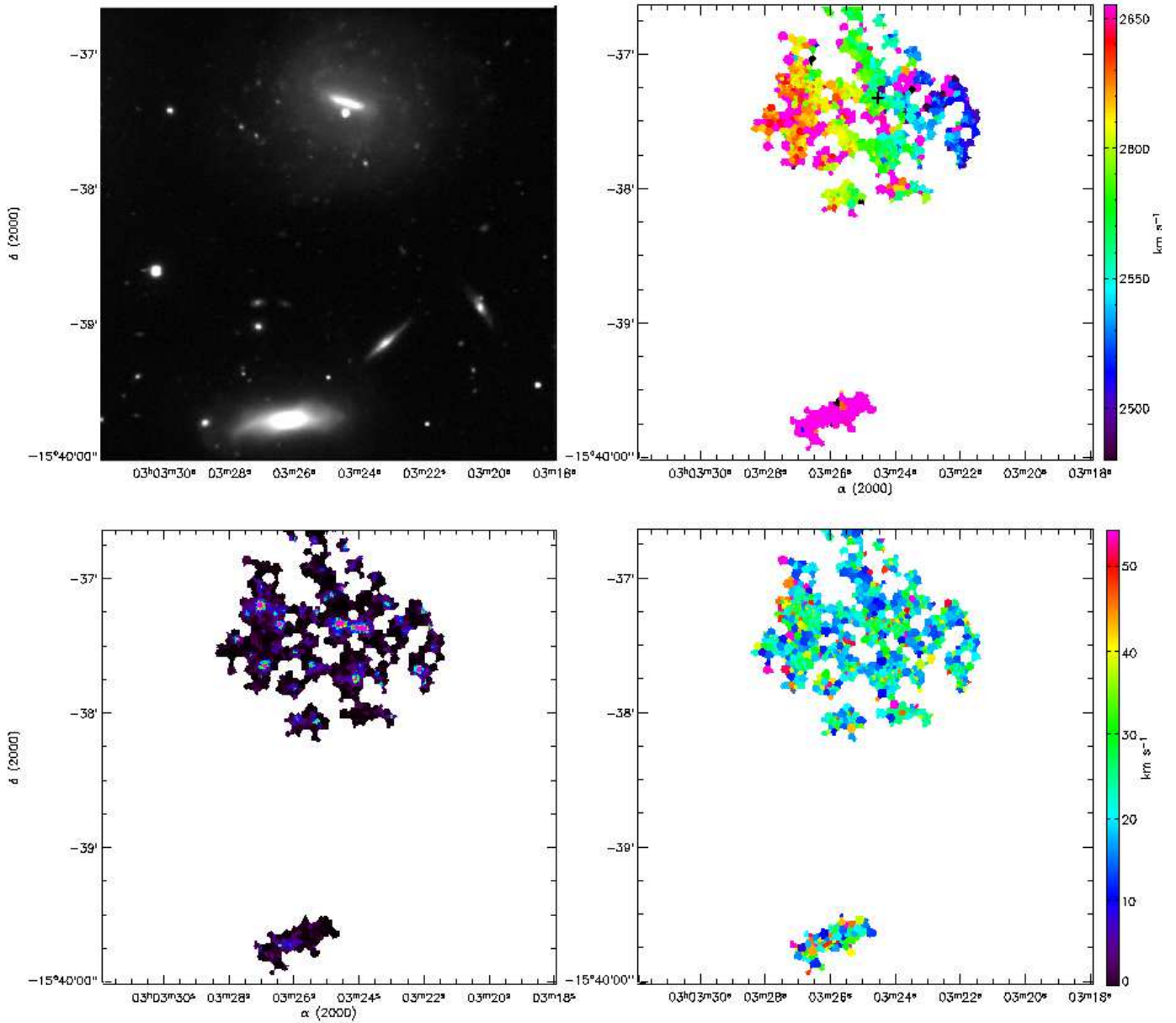}
\caption{HCG 22b,c. HCG 22b lies South of HCG 22c. Top left: R band image from CTIO. Top right: Velocity field. 
Bottom left: Monochromatic image. Bottom right: Velocity dispersion map.}
\label{maps_hcg22}
\end{figure*}
%\clearpage 

\begin{figure*}[t!]
\includegraphics[width=\textwidth]{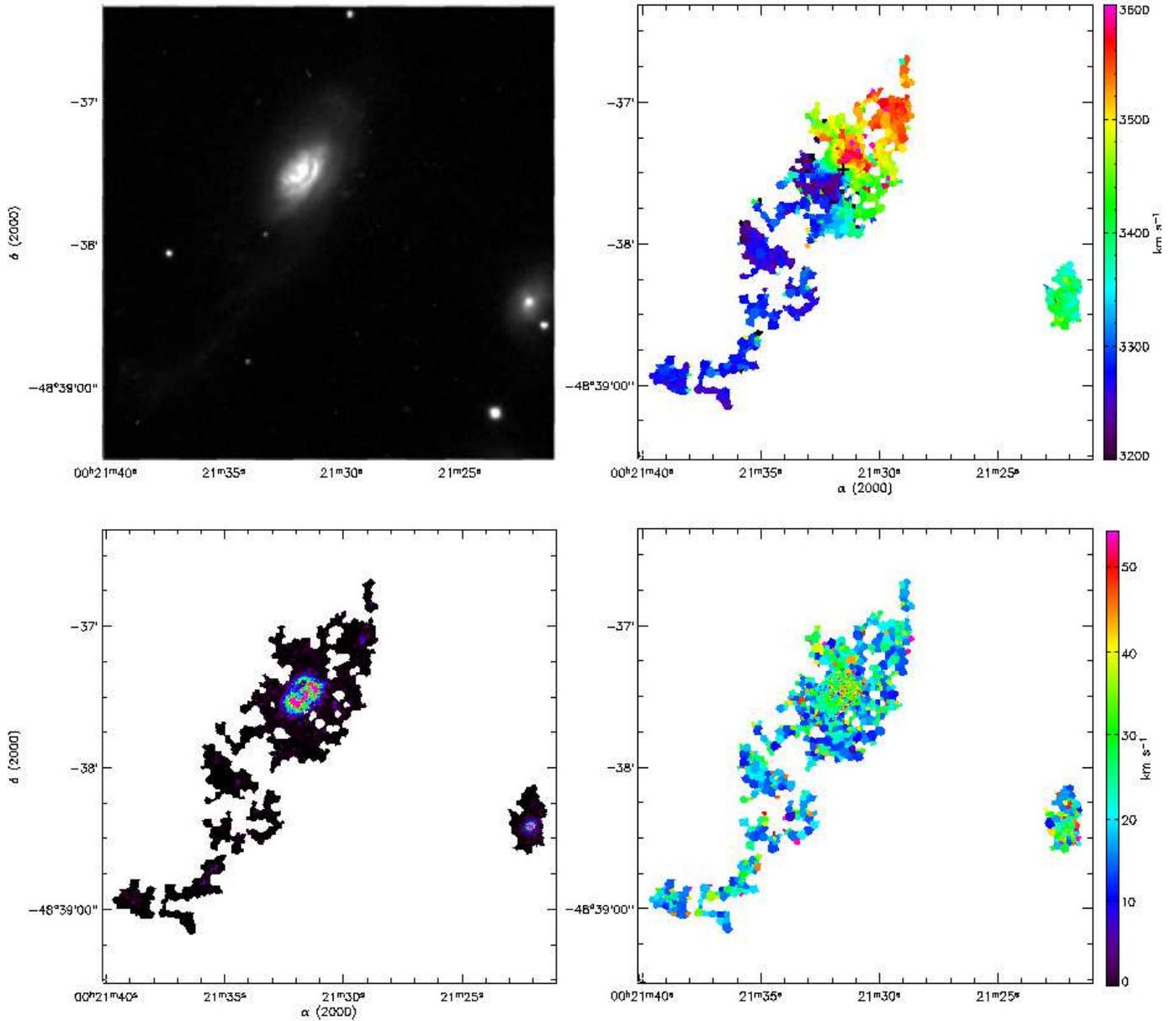}
\caption{NGC 92 and NGC 88. NGC 92 lies East of NGC 88. Top left: R band image from NTT. Top right: Velocity field. 
Bottom left: Monochromatic image. Bottom right: Velocity dispersion map.}
\label{maps_s01a}
\end{figure*}
%\clearpage

\begin{figure*}[t!]
\centering
\vspace{0.3cm}\includegraphics[scale=0.35]{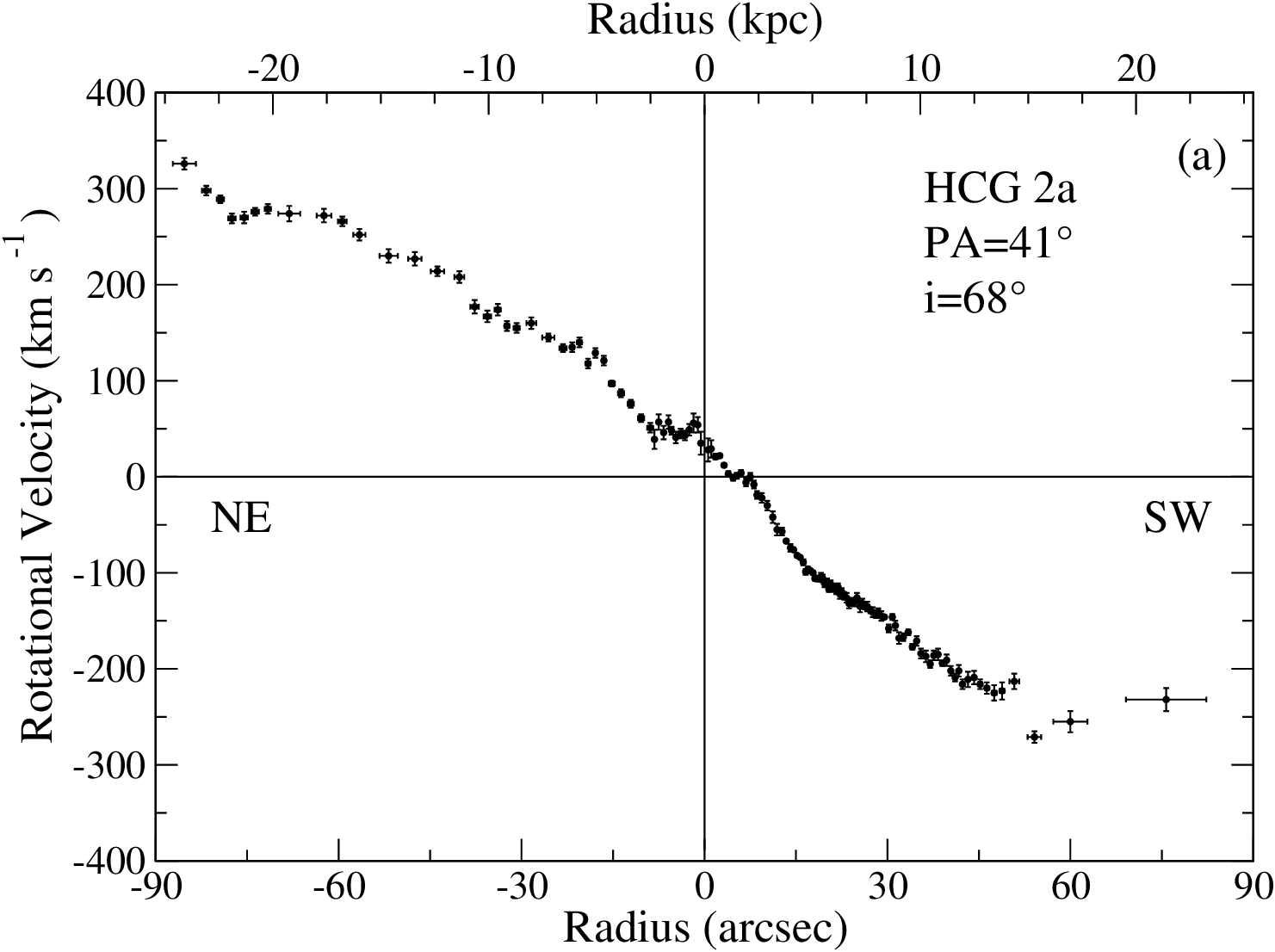}\hspace{0.05cm}\includegraphics[scale=0.35]{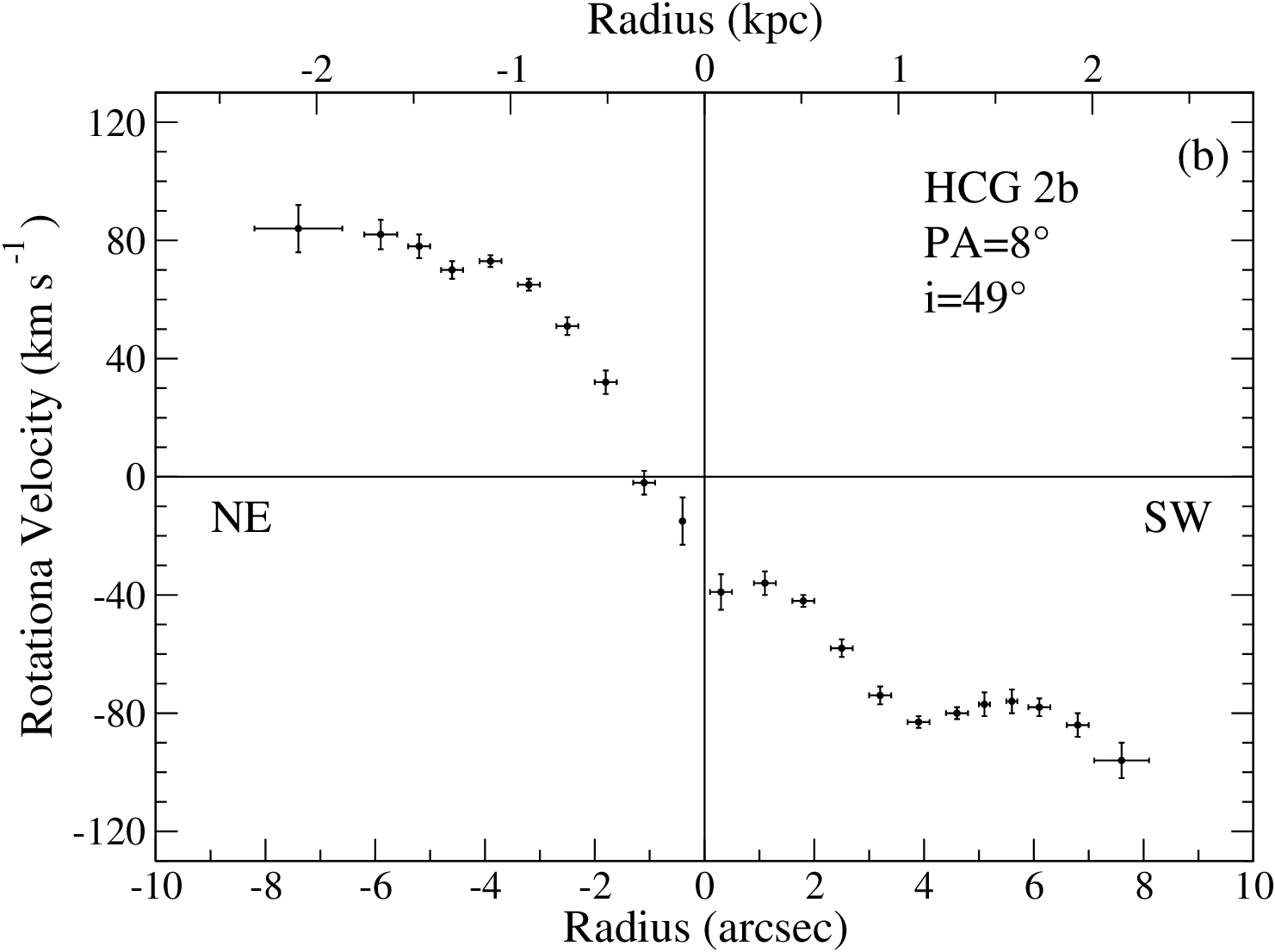} 
\label{hcg2a}
\end{figure*}
%0.36
\begin{figure*}[t!]
\centering
\includegraphics[scale=0.35]{fig_9.eps}\hspace{0.1cm}\includegraphics[scale=0.35]{fig_10.eps}
\label{hcg7a}
\end{figure*}

\begin{figure*}[t!]
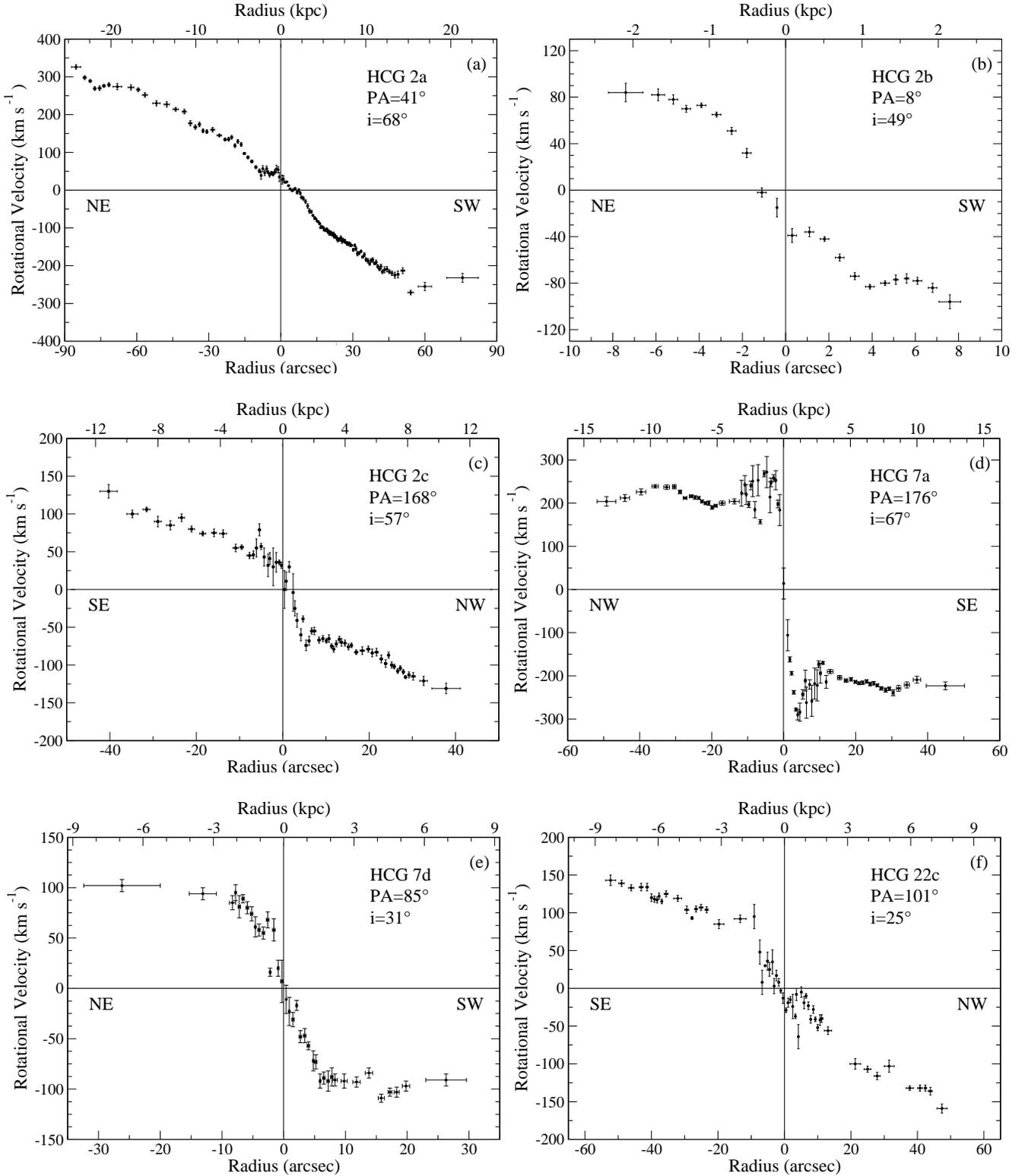

\centering
\hspace{-0.2cm}\includegraphics[scale=0.35]{fig_11.eps}\hspace{0.1cm}\includegraphics[scale=0.35]{fig_12.eps}
\caption{Rotation curves for individual galaxies in our sample. The parameters used to derive the rotation curves are given in each plot (PA:position angle, i:inclination). The adopted galaxy center is the morphological center. The horizontal error bars represent the $\pm$1$\sigma$ radius dispersion weighted by cos($\theta$), where $\theta$ is a polar coordinate in the plane of the galaxy.}
\label{hcg22c}
\end{figure*}

\subsection{Source Extraction and Photometry}

We searched for ultraviolet emitting regions in the vicinity of all seven targets (see section 4.2 for the corresponding figures). We also include HCG 92 and 
HCG 100 which were previously analyzed in Mendes de Oliveira et al. (2004) and de Mello et al.(2008a), 
using the following method. We generated catalogs using SExtractor (SE, Bertin \& Arnouts 1996)
in the FUV, NUV and R-band sky-subtracted images. The threshold for source 
detection was set to 1.5 sigma (DETECT$_{-}$THRESH=1.5) for a minimum area of 5 pixels (DETECT$_{-}$MINAREA=5). We recorded, 
for each source, the Right Ascention, Declination, MAG$_{-}$AUTO (AB system) and Petrosian radius. We used SE's Kron elliptical 
apertures to measure total magnitudes (MAG$_{-}$AUTO) in FUV, NUV and R-bands. Although  
MAG$_{-}$AUTO is often used to measure total magnitudes in galaxy surveys (e.g. Bell et al. 2004,  de Mello et al. 
2006, Zucca et al. 2006), high uncertainties might be expected at the faint magnitudes due to the assumption that 
the sky background has Gaussian random 
noise without source confusion (Brown et al. 2007). However, since we are comparing data taken with different 
resolutions and the fact that UV light does not necessarily peak at the same coordinate as the optical light, 
MAG$_{-}$AUTO performs better than the other SExtractor parameter choices, as long as a careful match between the different 
catalogs is performed (see de Mello et al. 2008a). 
We first matched the FUV and NUV catalogs using a 3$''$ radius and as a result we obtained the FUV and NUV parameters for each 
detected source. We then matched the UV catalog with the R-band catalog. Therefore, only objects with FUV detections 
were included in our final catalog. Due to the large \textit{GALEX} field of view (1$^{\circ}$.28 and 1$^{\circ}$.24 in FUV 
and NUV, respectively), we performed the source detection in four regions for each \textit{GALEX} image: one field centered on the
centroid of the group (on-group) and three other fields (each with $\sim$15$'$$\times$15$'$) distributed over the outskirts of the images (hereafter, Control Sample, CS).
The on-group field was defined within a region  of radius equal to two times the radius of the minimum circle that passes through the
center of each group member. This area was chosen because it usually encompasses the group and its HI debris, in case there is any.

Given that we used a different exposure time for each compact group observation we had to limit our analysis to the brightest completeness limit, which was that for the field of HCG 92, at a NUV magnitude of 22.2.
The colors FUV-NUV and FUV-R were estimated inside a  
fixed aperture of 4$''$ radius, centered on the centroid of the light distribution of each NUV band detection, 
using the task PHOT in IRAF.  Total magnitudes and fixed aperture magnitudes were corrected for galactic extinction, 
using A$_{FUV}$=E(B-V)$\times$8.29, A$_{NUV}$=E(B-V)$\times$8.18 (Seibert et al. 2005) and 
A$_{R}$=E(B-V)$\times$2.634 (Schlegel et al. 1998), following de Mello et al. (2008a).

\section{Results}

\subsection{Kinematic properties of the main galaxies}

\begin{figure}[t!]
\centering
\vspace{0.2cm}\includegraphics[scale=0.35]{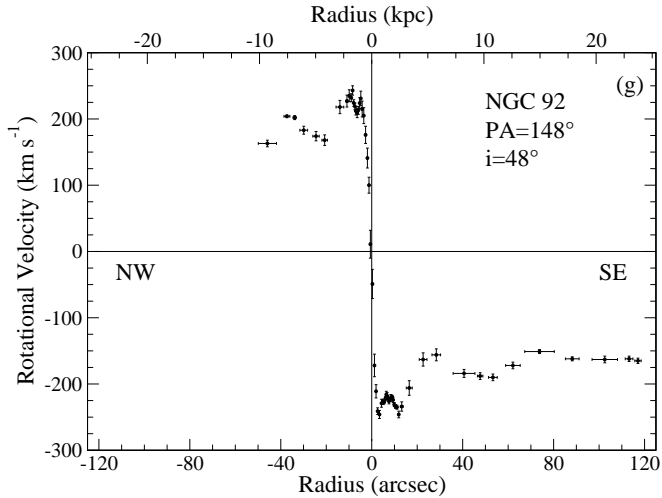}
\caption{Rotation curve for NGC 92. See also caption of Fig. 7.}
\label{ngc92}
\end{figure}

In this section, we analyze each individual galaxy based on our
optical R-band images, velocity fields, monochromatic images, velocity
dispersion maps (Figs. \ref{maps_hcg2ab} to \ref{maps_s01a}), and rotation
curves (Fig. \ref{hcg22c} and \ref{ngc92}), together with information
from the literature.  Note that the dispersion velocity maps were
corrected for instrumental broadening, measured using the calibration
cube. We estimated that the broadening is, on average, 24.8 km s$^{-1}$
accross the field.  In the last columns of Table \ref{table1} we list
the kinematic parameters obtained from the velocity fields. In columns
5 and 6 of Table \ref{table1} the systemic velocities taken from the
literature and the velocities resulting from our analysis are listed.

\subsubsection{HCG 2}

This compact group is formed of three galaxies with accordant redshifts.
According to the NED\footnote{http://nedwww.ipac.caltech.edu/} database and
the RC3
catalogue (de Vaucouleurs et al. 1991), the brightest galaxy
of the system, HCG 2a, is classified as a
late-type barred SBd galaxy. HCG 2b is classified as a cI (compact Irr)
by Hickson (1993) and it is the strongest FIR source in this group (Allam et
al. 1996).
HCG 2c is an SBc-type galaxy. Figure \ref{maps_hcg2ab} shows the velocity
field, the velocity dispersion, monochromatic and
continuum maps for HCG 2a,b.

HCG 2a shows a very disturbed velocity field as can be seen in Fig.
\ref{maps_hcg2ab}. No symmetry is observed with respect to the major axis and
moreover
one cannot see a constant velocity along the minor axis. The PA of the
velocity field
changes rapidly, by about 90$^{\circ}$ from the North to the South of the
galaxy. 
The velocity field shows the clear signature of a strong bar, 
almost vertical,
which is responsible for the PA variation and for the misalignment between the morphological and kinematic orientations. There is a misalignment of 30$^{\circ}$ between the bar and the kinematic minor axis.
This type of velocity field has been observed in Epinat et al. (2008b)  (e.g. UGC12754, UGC11407 and UGC1256). Given that the kinematic PA is so perturbed by the bar, it is better to take the morphological PA to draw the RC (see Epinat et al. 2008a).
The monochromatic map shows several HII regions. In the southern side, the regions are aligned along the spiral arm. Intense HII regions in the center of the object belong to the northern arm or to the bar. Towards the northern side of the galaxy only one bright region is visible, although other fainter HII regions are also seen around the same location. In the South side, we detected an extension of H$\alpha$ emission towards the West of this object. This signature was not reported in Vilchez et al. (1998). 
The dispersion map shows higher velocity dispersion in areas where HII regions are
present (Fig. \ref{maps_hcg2ab}).
The velocity dispersion within HII regions along the bar is 41 km s$^{-1}$
while outside these regions the dispersion is 25 km s$^{-1}$.
The velocity dispersion increases along the spiral arm, from 26 km s$^{-1}$
to 40 km s$^{-1}$.

The rotation curve of HCG 2a, shown in Fig. \ref{hcg22c}(a) 
is highly asymmetric, with the receding side reaching 326 km s$^{-1}$ and the approaching side 270 km s$^{-1}$, for a systemic velocity of 4347 km s$^{-1}$. The approaching side slope of the rotation curve is steeper than the receding side in
the SW side. For the adopted systemic velocity (4347 km s$^{-1}$), the rotation curve does not pass through the center of the diagram. 
However, both sides of the rotation curve match in the external region of the galaxy (outside 10\arcsec).
The center of the curve is particularly disturbed  due to the
strong bar. In this region the rotation curve is flat with a few oscillations
within 10 arcsec in both sides. The flatness of the curve is due to the orientation
of the bar with respect to the major axis of the galaxy (Hernandez et al.
2009, in preparation). The center of the bar coincides with the center of the
external ellipses of the continuum (for radii outside r=12$\arcsec$ in the minor axis and r=22$\arcsec$ in the major axis).

In the case of HCG 2b, we detected H$\alpha$ emission at its center
(within 2.8 kpc, also reported in Vilchez et al. 1998). The velocity field
(Fig. \ref{maps_hcg2ab}) seems to be regular. The kinematic position angle
differs from the morphological one (Mendes de Oliveira et al. 1992) by
20$^{\circ}$ at 2\textit{h$^{-1}$}kpc, however, at R$_{25}$, the optical
position angle is in good agreement with the kinematic one.  We noticed
that the continuum and monochromatic images peak at the same position.
%The receeding side shows a plateau outside 4" (1.3 kpc) while the approaching side shows two bumps placed between 0.5" (0.25 kpc) and 2" (0.5 kpc) and 4" (1 kpc) and 7" (2 kpc), respectively.  
The galaxy center displays a velocity dispersion of 47 km s$^{-1}$ while, in the outskirts the value is 34 km s$^{-1}$. We can also notice that a high value for the dispersion often corresponds to strong monochromatic
emission. The rotation curve of HCG 2b (Fig. \ref{hcg22c}b) shows a
disk-like rotator shape. For the adopted systemic velocity, both sides
match at radii larger than 2\arcsec, however, in the inner part of the
rotation curve, (inside 2\arcsec) the approaching side shows a plateau
and the two sides do not match. We detected two bumps in the
approaching side of the rotation curve, placed between 0$\arcsec$ and
4\arcsec and 4\arcsec and 7\arcsec, respectively.

The velocity field of HCG 2c is regular (see Fig. \ref{maps_hcg2c}). A signature of the bar is suspected in the central regions. The PA is almost constant along the major axis, staying around a value of
168$^{\circ}$. 
The monochromatic map shows several bright HII regions in the bar, in the center of the galaxy and across the northern arm. A tidal arm is identified in the monochromatic map, in a similar way to UGC 7861, as was reported by Epinat el al. (2008a). 
There are no bright HII regions in the south side of this object. 
The velocity dispersion map does not reveal any peculiarities. HCG 2c displays a more symmetric rotation curve than the other two members of this group, except in the central regions ($\leq$10$\arcsec$). The
velocity amplitude is $\sim$110 km s$^{-1}$, across 30$\arcsec$ (8 kpc). We also notice that the
continuum center is 2$\arcsec$ (0.5 kpc) away from the kinematic center,
towards the SE of the galaxy.

%In the Fabry-Perot analysis of the group HCG 2, we detect one source in the monochromatic image, located at the NW of HCG 2b, which is probably associated with optical counterparts in the R-band image (Fig. \ref{maps_hcg2ab}) at position RA=00:31:17.86, DEC=08:29:9.74.

The emission detected in the upper right corner of Fig. \ref{maps_hcg2ab} seems to be from a background galaxy, i.e., just an emission line falling in
the same velocity range as H$\alpha$ for HCG 2. Its morphology, as per inspection of the CTIO image of the group, seems to be that of a giant galaxy at a high 
redshift.

\subsubsection{HCG 7}

This group is formed by four objects, three late-type galaxies (Sb, SBc and SBc) and one early-type (SB0)
galaxy (Hickson 1993). HCG 7a and HCG 7c are strong
FIR sources (Allam et al. 1996, Verdes--Montenegro et al. 1998). Shimada et al. (2000) classified the nuclear activity in HCG 7a as an HII type.

HCG 7a shows H$\alpha$ emission in several knots through the arms (see Fig.\ref{maps_hcg7ad}) as was also noted
in Vilchez et al. (1998). The central region displays a peak in the monochromatic and continuum image. The velocity field of this galaxy is very peculiar, strong asymmetries around the major axis are observed on both sides of the galaxy.  Moreover, in the center (inside 3 kpc), strong non-circular motions are observed with the velocities reaching
high values of 290 km s$^{-1}$ in the rotation curve (see Fig. \ref{hcg22c}d). \
The velocity dispersion map for HCG7a displays a high value of
55 km s$^{-1}$ in the center of the galaxy, corresponding
to the central HII region visible in the monochromatic map. The other
HII regions of this galaxy do not show high velocity dispersions.
The bump in the central region of the rotation curve has already been 
found in other galaxies such as UGC 6118 (Epinat et al 2008a). The peak in the inner part corresponds to a region of high monochromatic emission and might be the result of a small bar, associated with the HII regions in the central parts. However, there is no previous mention of a stellar bar present for this galaxy. Both stellar and gaseous major axes appear to be aligned, probably 
due to the high inclination of the galaxy.

The SBc galaxy HCG 7d has been observed twice, the two observations are presented in Figs \ref{maps_hcg7ad} and \ref{maps_hcg7bd}.  This galaxy shows a strong H$\alpha$ emission at the end of its bar, visible in the
HST archival image and in several knots in their southwest arm.  
In the northeast side of this object, the intensity of H$\alpha$ emission is
lower than that in the south-west side. The monochromatic map of HCG 7d shown in Fig. \ref{maps_hcg7bd} displays emission in the east and west side of the galaxy, which is not detected in the monochromatic image of Fig. \ref{maps_hcg7ad} neither in the r-band image. We suspect that this signature is noise or residual emission from the OH subtraction.
The velocity field of HCG 7d clearly shows the presence of the bar (see Figure \ref{maps_hcg7bd}). The kinematic PA differs in 27$^{\circ}$ and 50$^{\circ}$ compared to the optical PA given in Mendes de Oliveira et al. (2003) and Hyperleda,
respectively. The velocity dispersion map does not show any particular
features. The rotation curve is quite regular, reaching a plateau of 100 km s$^{-1}$ from 7 (2 kpc) to 30$\arcsec$ (8 kpc). We conclude that this galaxy has a regular velocity field and rotation curve.

HCG 7b shows a weak H$\alpha$ emission in its center but the observations are truncated by the edge of the field of view. No information can be
extracted
from this velocity field.

Mendes de Oliveira et al. (2003) analyzed HCG 7c using Fabry-Perot data and
did not find peculiarities in its velocity field.

We note that the rotation curves for the galaxies in this group do not reach
the optical radius R$_{25}$, except for
HCG 7c. This could be
related to the fact that in HCG 7 about 80\% of the HI content of the member galaxies
is missing (Verdes-Montenegro et al. 2001). 

\subsubsection{HCG 22}

This group, composed of a triplet of galaxies with concordant redshifts, contains a bright 
elliptical and two spiral galaxies in its core (morphological types E2, Sa and SBcd according to Hickson 1993). Coziol et al. (1998a) classified the nuclear region 
of HCG 22a as a dwarf Seyfert 2 and Gallagher et al. (2008) did not find evidence for 
any excess mid-infrared emission in this object. Verdes--Montenegro et al. (1998) gave upper limits for the IR luminosities of HCG 22b and HCG 22c (log $L_{IR}$$<$8.55 $L_{\odot}$ and log $L_{IR}$=8.88 $L_{\odot}$ respectively). 
%and de Carvalho et al. (1997) showed that HCG 22b is an absorption-line galaxy and HCG 22c is an emission line-galaxy. 

The monochromatic map of HCG 22c (in Figure \ref{maps_hcg22}) shows several HII regions which belongs to the arms of 
this galaxy, showing a patchy appearance. Moreover, there is a faint H$\alpha$ emission 
in the intra-arm region. The velocity field seems to be regular, with a misalignment of about
30$^{\circ}$ between the kinematic and optical PA. The velocity dispersion map does not show any particular features.
The rotation curve displayed in Fig. \ref{hcg22c}(f) 
shows almost a linear shape, as
a solid-body rotator, particularly in the northwest part of the galaxy,
steadily rising from 0 to 125 km s$^{-1}$ out to 30\arcsec (5 kpc). From
30\arcsec (5 kpc) to 45\arcsec (7 kpc) the velocity rises from 125 km
s$^{-1}$ to 170 km s$^{-1}$. In the inner part of the rotation curve
(inside 15\arcsec), the approaching and receding sides do not match,
however, from 20 to 45\arcsec, both sides are in agreement.

\begin{figure}[t!]
\includegraphics[scale=0.47]{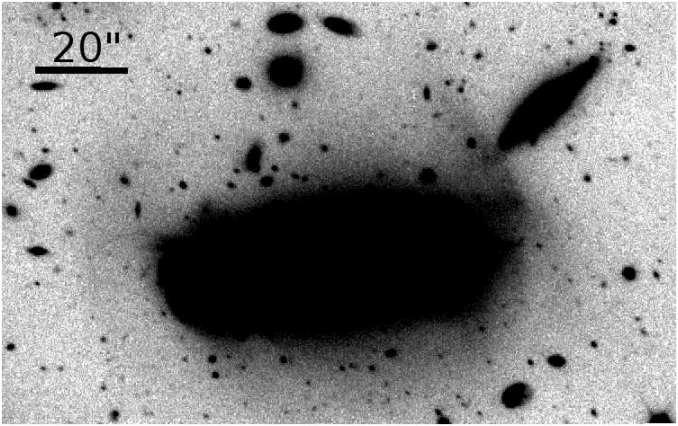}
\caption{R-band image of HCG 22b.}
\label{shellhcg22b}
\end{figure}

The early-type spiral HCG 22b only shows a weak H$\alpha$ emission in its 
central region. Therefore, the velocity field, which appears to be uniform, is poorly determined and it is not possible to obtain a rotation 
curve with the data in hands. 
HCG22b shows a high velocity dispersion equal to 48 km s$^{-1}$ 
in its center, which corresponds to the bright emission visible in 
the monochromatic map. HCG 22b exhibits extended shells clearly seen in the B and R-band images, as shown in Fig \ref{shellhcg22b}.

\subsubsection{NGC 92}

This galaxy is part of the Phoenix compact group formed by NGC 92, NGC 88 and NGC 89 and NGC 87 (Rose 1979). This group is also known as Rose 34 (Rose 1977) and  SCG 0018-4854 (Iovino et al. 2002). NGC 92 and NGC 88 are LINERs (Coziol et al. 2000) while NGC 89 is a Seyfert 2 galaxy (e.g. Pompei et al. 2007).

NGC 92 is classified as an Sa galaxy (Prugniel et al. 1998). 
It shows a prominent tidal tail in the South East direction clearly visible also in the H$\alpha$ image of Temporin et al. (2005). The velocity field presented in 
Fig. \ref{maps_s01a} shows a disk-like rotation with a large radial velocity range (from 3641 km s$^{-1}$ to 3183 km s$^{-1}$). The position angle of the 
velocity field changes from 8$^{\circ}$ to 28$^{\circ}$ over a 40 arcsec radius. Even with this strong  change in the position angle, 
the velocity field is fairly regular. The very center of the velocity field clearly shows double components. 
The second component is better defined in the blue part of the velocity field than in the red part, where it  
fades. Along the tail, this second component appears at the tip of the tail, but it is not possible 
to follow it all across the tail. Such double profiles have been observed in other asymmetric galaxies (e.g. UGC 3334, GHASP sample) and a tidal arm with an almost constant radial velocity has also been observed (e.g. UGC 3429, GHASP sample). The monochromatic map (Fig. \ref{maps_s01a}) shows a very bright emission in the galaxy center. Much lower 
intensity emission regions are located in the northern region and across the tidal tail. These coincide with blobs also detected in the NUV-band image from \textit{GALEX} (see Fig. \ref{ngc92nuv}). The central emission of NGC 92 shows a geometry consistent with spiral arms. 
%with three very intense regions. 
There are a few hot spots along the tail also consistent with the \textit{GALEX} image (see Fig. \ref{ngc92nuv}). 
The maximum velocity dispersion shown is $\sim$ 70 km s$^{-1}$ 
but looking at the profiles it 
is obvious that several components are present. 
%Regions with high emission show moderate velocity dispersions ($\sim$35
%km s$^{-1}$). Away from the high emission region (center) and regions
%in the tail show lower
%velocity dispersions ($\sim$15 km s$^{-1}$).
The dispersion map shows a higher velocity dispersion in the
center of the galaxy than in its outskirts and tidal tail.  In the
center the average velocity dispersion is 45 km s$^{-1}$. Regions 1,
2 and 3, detected with GALEX, have a lower velocity dispersion, lower than 
25 km s$^{-1}$. %Both sides of the rotation curve (Fig. \ref{ngc92}) show a plateau from 2.5$\arcsec$ (0.5 kpc) to 18$\arcsec$ (3.5 kpc). 
The receding side reaches a rotational velocity of 243 km s$^{-1}$ and the
approaching side reaches 246 km s$^{-1}$. The long tidal tail appears
as an extension of the approaching side of the rotation curve, with a
constant velocity of 160 km s$^{-1}$ from 70 to 120\arcsec. As HCG 7a, the rotation curve of NGC 92 presents a bump in velocity within 20\arcsec (5kpc) that could be interpreted as a signature of non circular motion.

We also detected H$\alpha$ emission at the center of NGC 88. However, we did not attempt to derive any rotation curve because this object is located at the edge of the image. However, we note that the velocity field appears uniform.

\subsection{Field Density of UV emitting regions}

\begin{figure}[ht!]
\includegraphics[scale=0.47]{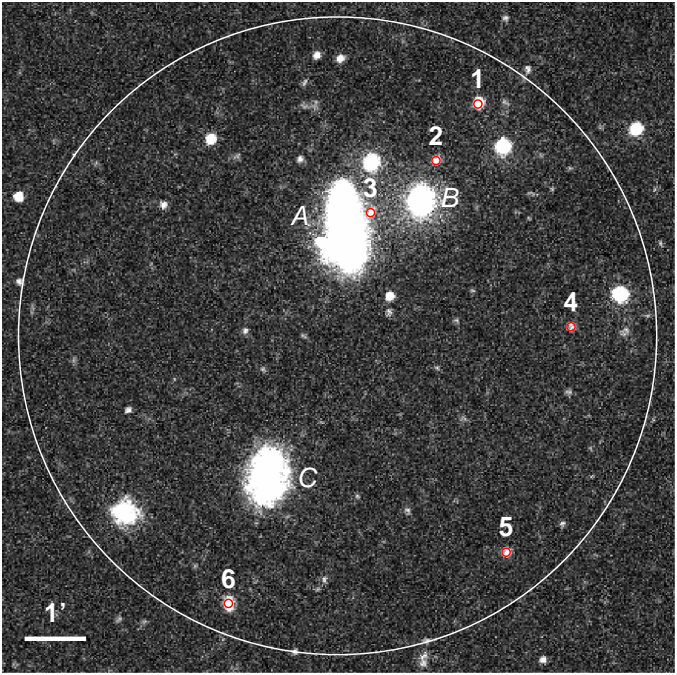}
\caption{NUV band image for HCG 2. The smaller circles show the detected regions. The big circle has a radius of twice the minimum circle that contains the galaxies members of the group, which are named with capital letters.}
\label{hcg2nuv}
\end{figure}

\begin{figure}[ht!]
\includegraphics[scale=0.47]{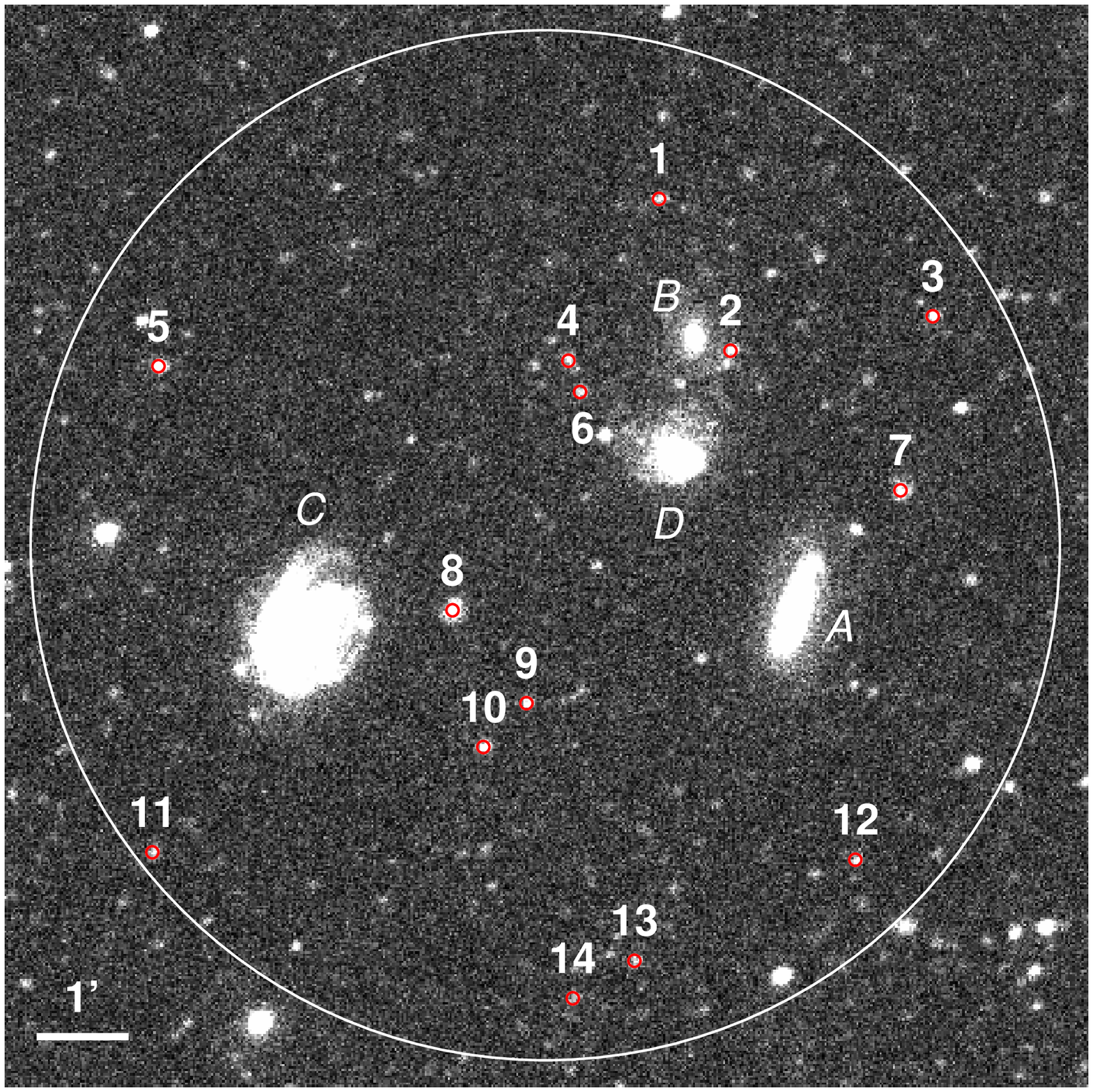}
\caption{NUV band image for HCG 7. See caption of Figure \ref{hcg2nuv}.}
\label{hcg7nuv}
\end{figure}

\begin{figure}[ht!]
\includegraphics[scale=0.47]{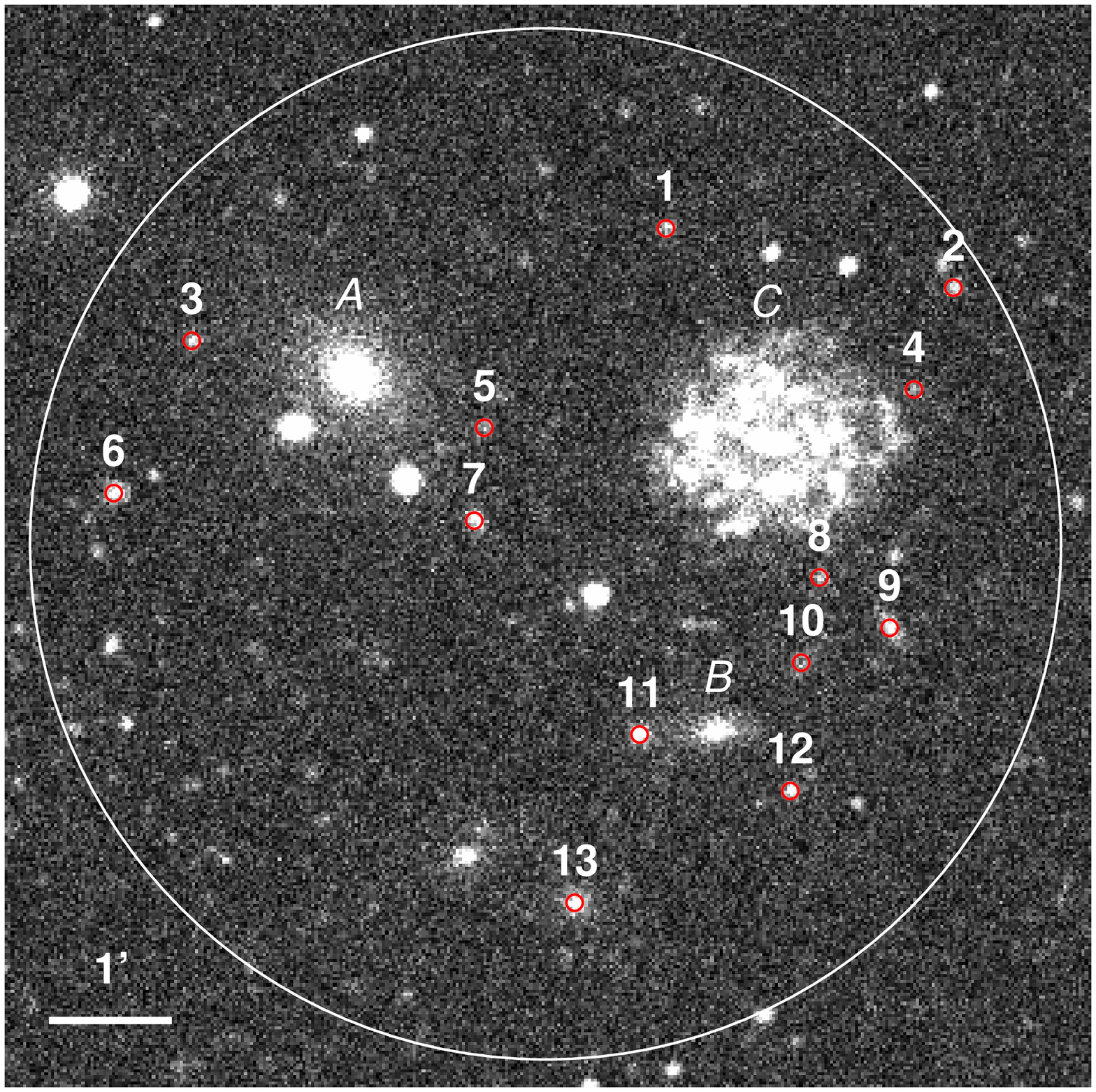}
\caption{NUV band image for HCG 22. See caption of Figure \ref{hcg2nuv}.}
\label{hcg22nuv}
\end{figure}

\begin{figure}[ht!]
\includegraphics[scale=0.47]{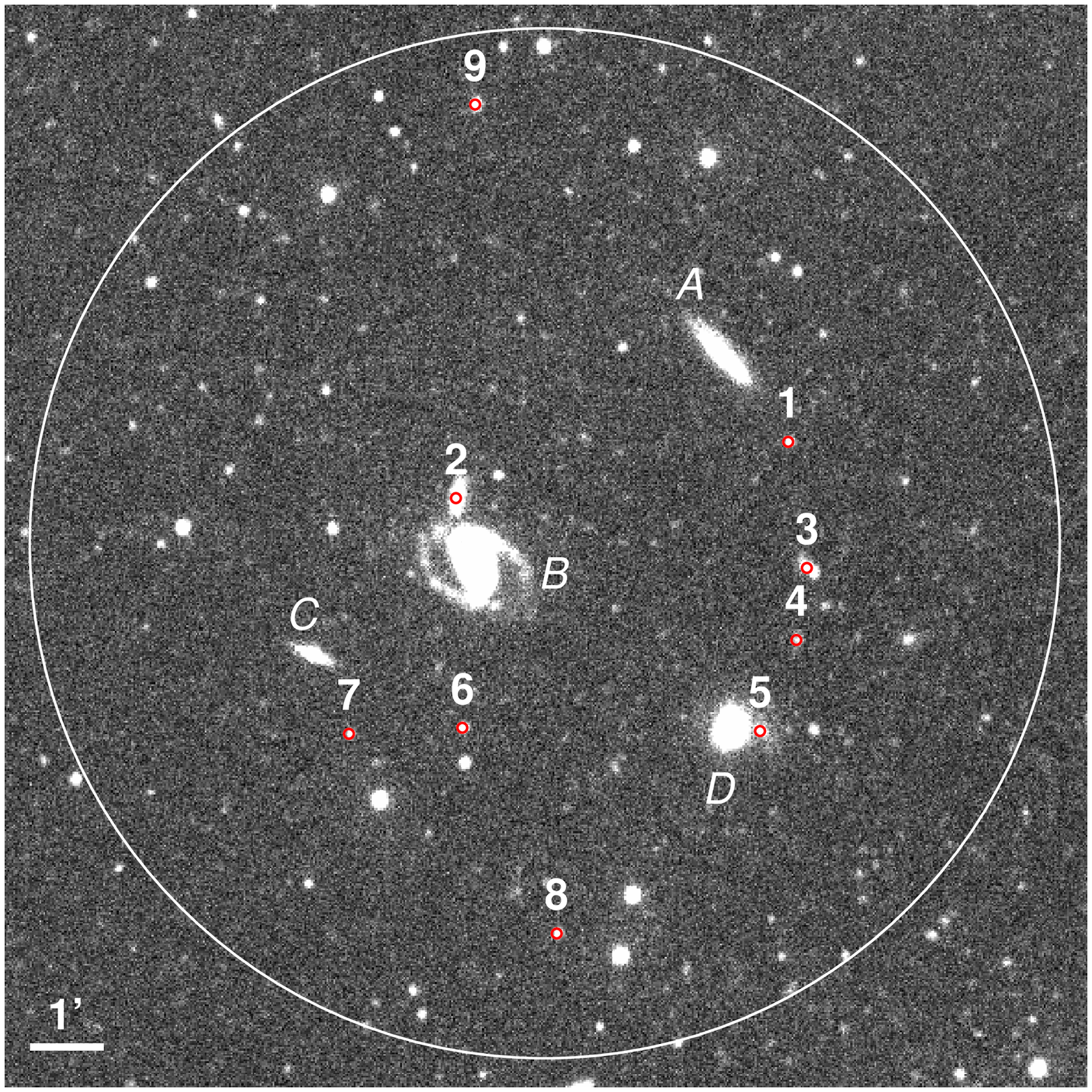}
\caption{NUV band image for HCG 23. See caption of Figure \ref{hcg2nuv}.}
\label{hcg23nuv}
\end{figure}

\begin{figure}[ht!]
\includegraphics[scale=0.47]{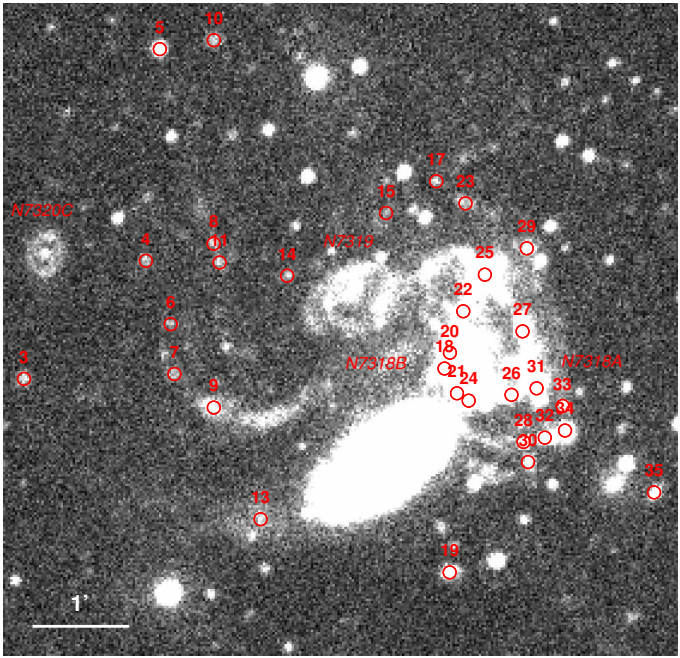}
\caption{Zoom of the central region of HCG 92 (NUV band image). See caption of Figure \ref{hcg2nuv}. Galaxy names were taken from Mendes de Oliveira et al. 2004.}
\label{hcg92nuv}
\end{figure}

\begin{figure}[ht!]
\includegraphics[scale=0.47]{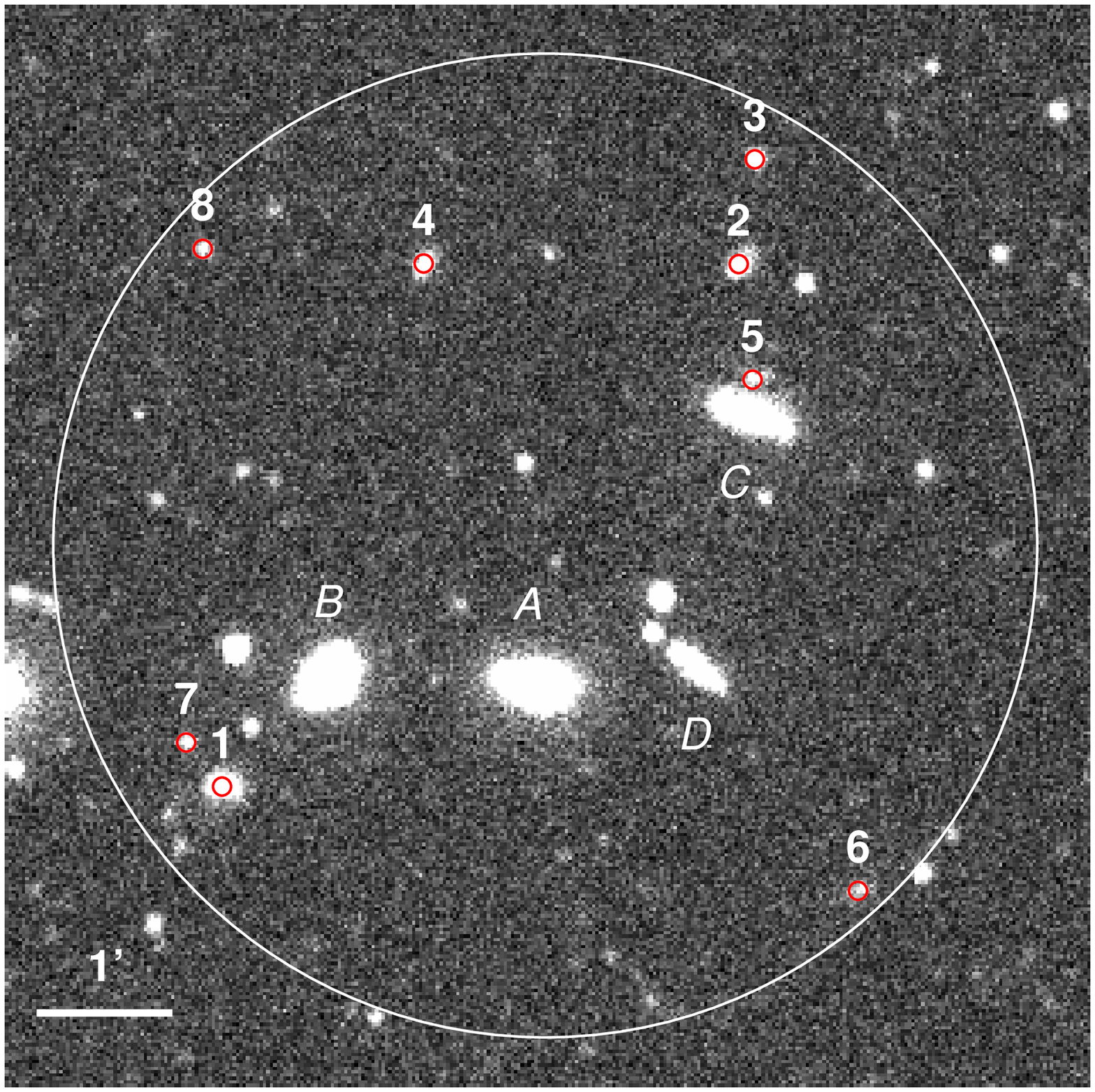}
\caption{NUV band image for HCG 100. See caption of Figure \ref{hcg2nuv}.}
\label{hcg100nuv}
\end{figure}

\begin{figure}[ht!]
\includegraphics[scale=0.47]{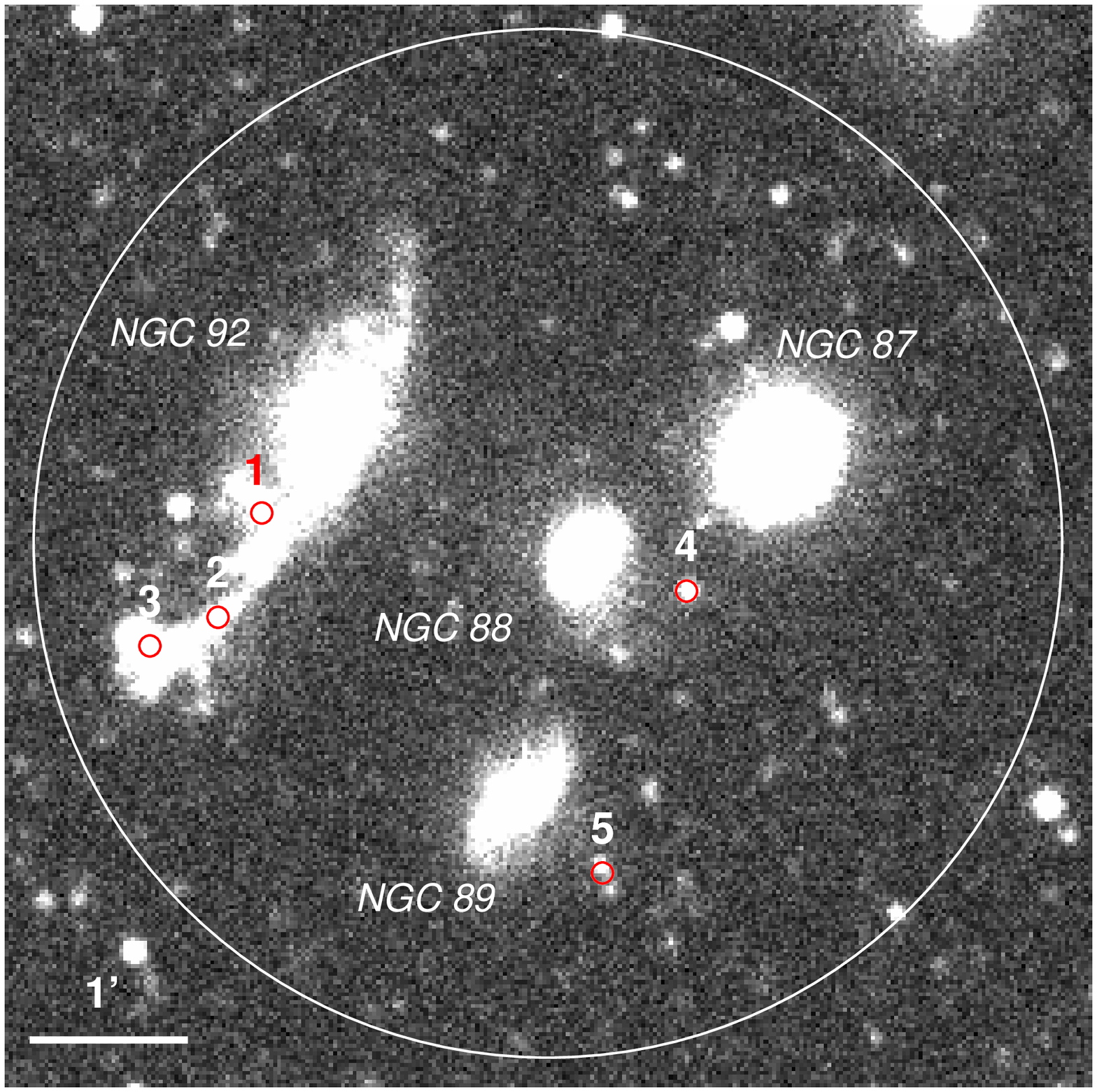}
\caption{NUV band image for NGC 92. See caption of Figure \ref{hcg2nuv}.}
\label{ngc92nuv}
\end{figure}

In Figures \ref{hcg2nuv} to \ref{ngc92nuv} we show the \textit{GALEX}
NUV images for each compact group where sources are numbered in circles
of 4$''$ radius.  We cannot exclude the possibility that these
regions are not related to the groups, given that we do not have redshift
information for most of them.  However, a similar search by de Mello et
al. (2008a) using GALEX images of HCG 100 was able to detect two TDGs
within the HI tail (which were spectroscopically confirmed by Urrutia et
al. 2009). Moreover, de Mello et al. (2008b), using GALEX images, have
detected star-forming regions in the vicinity of the HI bridge between
M81 and M82. Therefore, these two multiwavelength studies show that this
approach is successful in identifying at least some star-forming regions
in the intragroup medium. Determination of the membership and true nature
of these regions will have to await spectroscopy with large telescopes.

In order to evaluate whether our targets contain an excess of ultraviolet
emitting regions in their vicinity, we have calculated the density of
objects in the compact group field and in the control sample (CS).
The field density was defined as the number of regions inside a
circle of two group radii (see section 3.2) divided by the area of
that field after exclusion of the areas occupied by the
galaxies, as described below). Regions detected 
inside the optical contours of group members (within R$_{25}$) were eliminated,
since we are searching for star formation outside galaxies. 
This was done by discarding all regions inside an ellipse with major
and minor axes \textit{a} and \textit{b}, which correspond to the length of
the major and minor axes of the $\mu$$_{B}$=25 mag arc$^{-2}$ 
isophote given in Hickson (1993). For galaxies HCG 92B and D, we found 
necessary to recompute the region within the R$_{25}$ isophote using the 
B image presented in Mendes de Oliveira et al. (2001), given the strong isophote overlap between these two systems.  
We
note that three star-forming regions located over the tidal tail of
NGC 92 were included in our analysis, given that these objects are located
at projected distances outside the main body of this galaxy.
We have disconsidered the area
of each galaxy when estimating the total area of the group. No excess of
ultraviolet emitting sources is detected in HCG 2, HCG 7, HCG 23 and NGC
92, when the groups are compared with the CS fields. Table \ref{table2} shows
the spatial densities of objects around each compact group and in three
control fields.  HCG 92 contains the highest field density (0.28$\pm$0.05
objects arcmin$^{-2}$) in our sample. Most of the regions detected in
this group are located in the tidal tails of NGC 7318A, NGC 7318B and
NGC 7319. We detected UV emission in all intergalactic HII regions found
by Mendes de Oliveira et al. (2004), confirming the results by Xu et
al. (2005). Besides HCG 92, HCG 22 has also an excess of ultraviolet
sources compared with their control sample. In this group, some of
the UV detected regions show 8 $\mu$m emission (Johnson et al. 2007).
Although the field density of HCG 100 is slightly higher than in the
CS fields, the error bars are large.  Therefore, we have not found any
significant excess of ultraviolet emitting regions in the intragroup field
of HCG 100, when we compared to the field density of the control sample.
We note that the background counts are solely based on the observed 
counts in the control fields, not taking into account existing redshift
information for individual regions, given that very few of them have 
known redshifts.

\begin{table*}
\caption{Field Density $\Sigma$ (\#objects arcmin$^{-2}$)}
\label{table2}      
\centering          
\begin{tabular}{cccccccc}     % 7 columns 
\hline\hline 
Compact Group & $\Sigma$ Compact Group & $\Sigma$ Control sample 1 & $\Sigma$ Control sample 2 & $\Sigma$ Control sample 3\\
\hline
HCG 2    & 0.07$\pm$0.03 & 0.07$\pm$0.02 & 0.05$\pm$0.02 & 0.16$\pm$0.03  \\
HCG 7    & 0.15$\pm$0.04 & 0.16$\pm$0.03 & 0.20$\pm$0.03 & 0.17$\pm$0.03  \\
HCG 22   & 0.25$\pm$0.07 & 0.12$\pm$0.03 & 0.11$\pm$0.03 & 0.13$\pm$0.03 \\
HCG 23   & 0.06$\pm$0.02 & 0.08$\pm$0.02 & 0.13$\pm$0.03 & 0.11$\pm$0.03  \\
HCG 92   & 0.28$\pm$0.05 & 0.05$\pm$0.02 & 0.08$\pm$0.02 & 0.08$\pm$0.02 \\
HCG 100  & 0.20$\pm$0.07 & 0.09$\pm$0.02 & 0.08$\pm$0.02 & 0.14$\pm$0.03  \\
NGC 92   & 0.16$\pm$0.07 & 0.23$\pm$0.03 & 0.20$\pm$0.03 & 0.12$\pm$0.02 \\
\hline                  
\end{tabular}
\end{table*}

\subsection{FUV-NUV color distribution}

Since we are interested in young regions, we conservatively focused this analysis on 
objects bluer than FUV--NUV=1.
We chose this limit based on the work of Gil de Paz et al. (2007) where it is
reported that colors bluer than FUV--NUV=0.9 are typical of spiral and 
late-type galaxies. 
Objects with colors redder than 
FUV--NUV=1 (see Fig. \ref{FUV_R_FUV_NUV_2} and \ref{NUV_R_FUV_NUV}) account only for 6\% of the total intragroup regions, 5\% being stars (see \S 4.4). In the case of the control sample, 
objects with colors redder than FUV--NUV=1 correspond to 6\% of the total. Tables \ref{table3} to \ref{table9} show the FUV--NUV colors for all detected regions in the compact group fields.

\begin{figure*}[ht!]
\includegraphics[width=\textwidth]{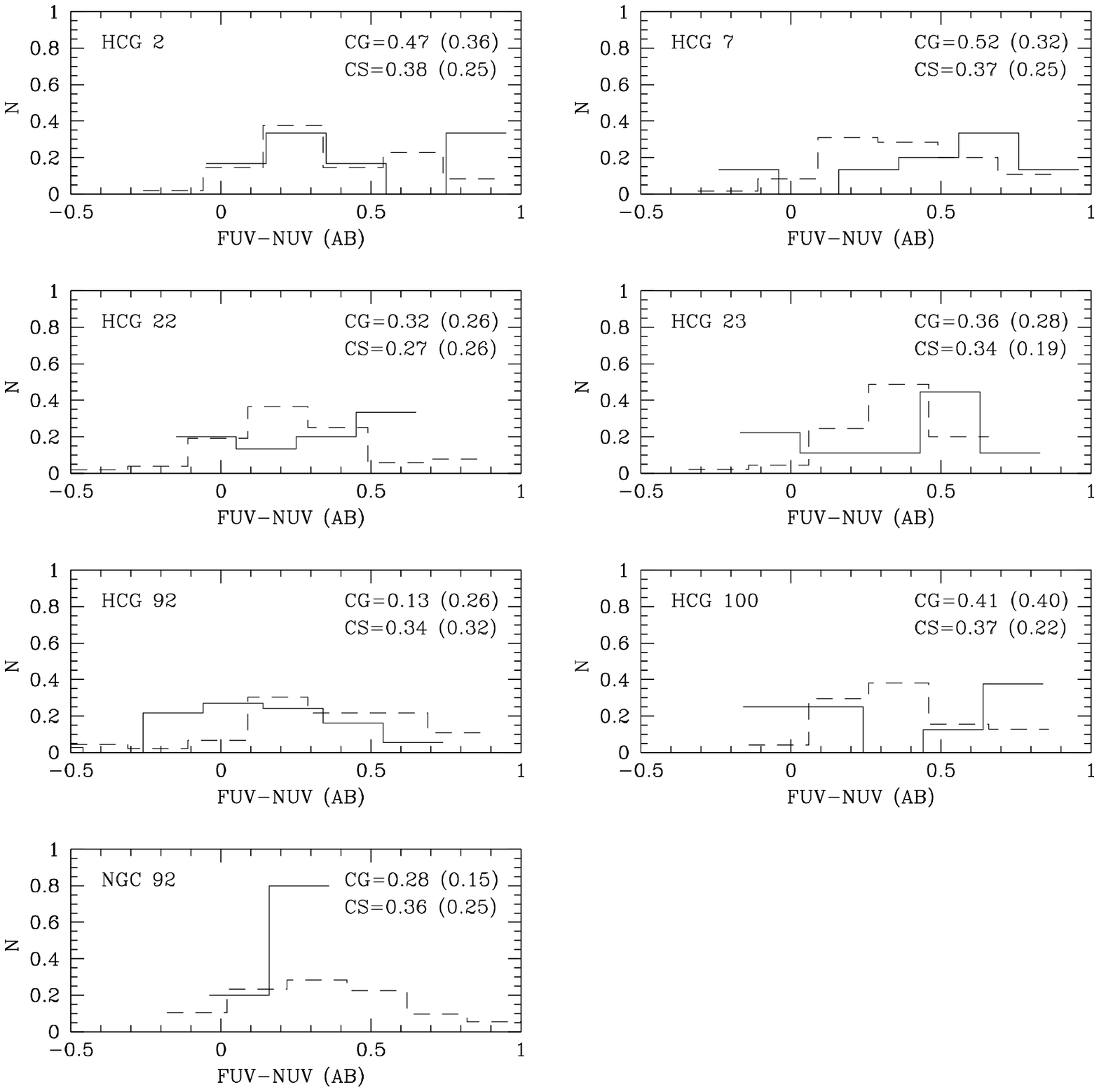}
\caption{Normalized histograms showing FUV-NUV distributions. Solid line: Intragroup regions. Dashed line: Control sample. 
Averages for each FUV-NUV color distribution are shown as CG, for Compact group and CS, for Control sample. Values for the standard deviation are indicated in parentheses. The error bar in the x-axis is estimated to be 0.2 mag.}
\label{allgroups_CS_05}
\end{figure*}

The FUV--NUV color distributions of the intragroup sources together with their respective CSs 
are shown in Figure \ref{allgroups_CS_05}. We found that the CS color distributions are similar 
among themselves with the exception of that for HCG 23. We also found that only the intragroup regions of 
HCG 92 and HCG 100 show a blue FUV--NUV color distribution, as compared to the CS. HCG 92 has the bluest FUV--NUV average 
in this sample, FUV--NUV=0.13$\pm$0.26. Although the average in FUV--NUV color for HCG 100 is not among the bluest in this sample, FUV--NUV=0.41$\pm$0.40, the color distribution shows a bimodal appearance, with a blue peak with mean color of zero.

The intragroup color distribution is similar for HCG 7 and HCG 22. In both cases the intragroup FUV--NUV average is redder than their CS, as shown in 
Fig. \ref{allgroups_CS_05}. From the same Figure, HCG 2 shows a bimodal distribution, with a peak in the 
red side of the FUV--NUV color and distribution similar to the control sample. In the case of HCG 23, the 
intragroup distribution shows no special signature compared with the control sample distribution. NGC 92 
presents a single peak in the FUV--NUV distribution, with an average color equal to 0.28$\pm$0.15.
We note that NGC 92 contains only five detected intragroup regions, the lowest number in our sample.

HCG 7 has the reddest FUV--NUV average of all groups, 
0.52$\pm$0.32. Of a total of 14 regions, 4 regions (29\% of the intragroup sample) 
present colors bluer than the average FUV-NUV color for their control sample and 3 of them are located close to HCG 7c.  
Region \#5 is the bluest object in this compact group and shows stellar characteristics in the SDSS image. This 
object is 57.9 kpc away from HCG 7c. 

The distribution of colors of intragroup objects in HCG 2 has an average FUV-NUV color of 0.47$\pm$0.38. We found 3 regions (50\% of the intragroup sample) bluer than the average FUV--NUV of their control sample: region \#4, \#5 and \#6, located  of 58.2 kpc, 75.9 kpc and 39.2 kpc away from the closest galaxy (HCG 2b, HCG 2c and HCG 2c respectively). Those objects show an extended morphology in our R-band image. Regions \#2 and \#3  seem to be associated with the regions identified in the monochromatic map of HCG 2, as we noted in Session 4.1.1.

In the case of HCG 22, the objects in the intragroup medium have an average FUV-NUV color of 0.32$\pm$0.26 and this group contains 6 regions 
(46\% of the intragroup sample) bluer than the average color of the control sample counterparts.

For HCG 23, the FUV-NUV average is 0.36$\pm$0.28, with 3 regions (33\% of the intragroup sample) being bluer than their control sample.

As we noted before, HCG 92 has the bluest FUV--NUV average and distribution. In this group, 
78\% of the intragroup candidates (28 regions) are bluer than the value for the FUV--NUV control sample average, 
mainly placed in the tidal tails, as has been reported in Mendes de Oliveira et al. (2001). 
The regions in common between that work and ours are indicated in the second
column of Table 7.

In HCG 100, we found 4 regions (50\% of the intragroup sample) bluer than the average values of the CS FUV--NUV: regions \#4 (\#4), \#6 (\#6), \#7 (\#13) and \#8 (\#14).
Note that our identification in Table \ref{table8} is not the same as in de Mello et al. (2008a), the latter are given in parenthesis.
 
In NGC 92, 60\% of the regions are bluer than the average color of the regions of the control sample. Interestingly, the bluest regions in this 
group, \#1, \#2 and \#3, are located in the tidal tail of NGC 92, where it is possible to note a gradient in the 
FUV-NUV color along the tail, with the bluest region located at the end of the tidal tail. Similar effects were 
reported by Hibbard et al. (2005) for NGC 4038/39.

\begin{table*}
\begin{minipage}[t]{\textwidth}
\caption{Observed and Derived Properties for HCG 2 Regions}
\label{table3}
\centering
\renewcommand{\footnoterule}{}  % to avoid a line before footnotes
\begin{tabular}{ccccccc}
\hline \hline
ID & R.A.& DEC. & FUV & NUV & FUV-NUV\footnote{Using a fixed apertures of 4\arcsec.} & Age\footnote{Age from FUV-NUV using Thilker et al. (2007).}\\
~ &(2000)&(2000) & ~& ~ & ~ &(Myr) \\
\hline
1 & 7.81289 & 8.50144 & 20.04 $\pm$ 0.04 & 19.04 $\pm$ 0.02 & 0.85 $\pm$ 0.05 &        \\
2 & 7.82440 & 8.48604 & 21.46 $\pm$ 0.10 & 20.67 $\pm$ 0.06 & 0.49 $\pm$ 0.11 &       \\
3 & 7.84238 & 8.47191 & 19.58 $\pm$ 0.05 & 18.73 $\pm$ 0.02 & 0.91 $\pm$ 0.12 &       \\
4 & 7.78734 & 8.44088 & 21.99 $\pm$ 0.14 & 21.40 $\pm$ 0.10 & 0.33 $\pm$ 0.14 &       \\
5 & 7.80517 & 8.37961 & 20.99 $\pm$ 0.07 & 20.85 $\pm$ 0.06 & 0.05 $\pm$ 0.09 & 40.8	  \\
6 & 7.88140 & 8.36569 & 20.00 $\pm$ 0.04 & 19.72 $\pm$ 0.03 & 0.16 $\pm$ 0.07 &         \\
\hline
\end{tabular}
\end{minipage}
\end{table*}

\begin{table*}
\begin{minipage}[t]{\textwidth}
\caption{Observed and Derived Properties for HCG 7 Regions}
\label{table4}
\centering
\renewcommand{\footnoterule}{}  % to avoid a line before footnotes
\begin{tabular}{ccccccc}
\hline \hline
ID & R.A.& DEC. & FUV & NUV & FUV-NUV\footnote{Using a fixed apertures of 4\arcsec.} & Age\footnote{Age from FUV-NUV using Thilker et al. (2007).}\\
~ &(2000)&(2000) & ~& ~ & ~ &(Myr) \\
\hline
1 &  9.83047  &  0.93806 & 21.86 $\pm$ 0.10 & 21.41 $\pm$ 0.09 & 0.36  $\pm$ 0.13 &	  \\
2 &  9.81752  &  0.91055 & 21.83 $\pm$ 0.11 & 20.74 $\pm$ 0.05 & 0.90  $\pm$ 0.13 &	  \\
3 &  9.78083  &  0.91679 & 22.10 $\pm$ 0.11 & 21.09 $\pm$ 0.06 & 0.70  $\pm$ 0.12 &	  \\
4 &  9.84691  &  0.90873 & 21.81 $\pm$ 0.10 & 20.95 $\pm$ 0.07 & 0.65  $\pm$ 0.14 &	  \\
5 &  9.92126  &  0.90774 & 20.22 $\pm$ 0.04 & 20.23 $\pm$ 0.03 & -0.14 $\pm$ 0.05 &   3.6      \\
6 &  9.84480  &  0.90307 & 22.48 $\pm$ 0.13 & 21.18 $\pm$ 0.08 & 0.93  $\pm$ 0.16 &	  \\
7 &  9.78668  &  0.88524 & 21.38 $\pm$ 0.08 & 20.35 $\pm$ 0.04 & 0.71  $\pm$ 0.11 &	  \\
8 &  9.86793  &  0.86351 & 20.29 $\pm$ 0.05 & 19.79 $\pm$ 0.03 & 0.49  $\pm$ 0.08 &	  \\
9 & 9.85452   & 0.84667  & 22.33 $\pm$ 0.12 & 21.33 $\pm$ 0.06 & 0.76  $\pm$ 0.13 &	  \\
10 & 9.86234   & 0.83872  & 21.24 $\pm$ 0.07 & 20.89 $\pm$ 0.05 & 0.27  $\pm$ 0.09 &	  \\
11 & 9.92238   & 0.81964  & 21.82 $\pm$ 0.09 & 21.76 $\pm$ 0.08 & -0.05 $\pm$ 0.12 &   12.2    \\
12 & 9.79487   & 0.81829  & 22.39 $\pm$ 0.13 & 21.61 $\pm$ 0.09 & 0.43  $\pm$ 0.14 &	  \\
13 & 9.83494   & 0.79998  & 22.46 $\pm$ 0.13 & 21.55 $\pm$ 0.09 & 0.75  $\pm$ 0.16 &	  \\
14 & 9.84611   & 0.79318  & 22.46 $\pm$ 0.14 & 21.65 $\pm$ 0.11 & 0.46  $\pm$ 0.17 &	  \\
\hline
\end{tabular}
\end{minipage}
\end{table*}

\begin{table*}
\begin{minipage}[t]{\textwidth}
\caption{Observed and Derived Properties for HCG 22 Regions}
\label{table5}
\centering
\renewcommand{\footnoterule}{}  % to avoid a line before footnotes
\begin{tabular}{cccccccc}
\hline \hline
ID & R.A.& DEC. & FUV & NUV & FUV-NUV\footnote{Using a fixed apertures of 4\arcsec.} & Age\footnote{Age from FUV-NUV using Thilker et al. (2007).}& Velocity\footnote{Velocity taken from de Carvalho et al. (1997) and Hickson et al. (1992) for object 9 and 13 respectively.}\\
~ &(2000)&(2000) & ~& ~ & ~ &(Myr) &(km s$^{-1}$)\\
\hline
1  & 45.86570  &  -15.59343 & 22.04 $\pm$ 0.19 & 21.45 $\pm$ 0.09 & 0.57  $\pm$ 0.20 &    &    \\
2  & 45.82517  &  -15.60148 & 21.79 $\pm$ 0.15 & 21.39 $\pm$ 0.09 & 0.10  $\pm$ 0.16 &   60.8 &     \\
3  & 45.93254  &  -15.60869 & 21.57 $\pm$ 0.14 & 21.31 $\pm$ 0.09 & -0.05 $\pm$ 0.16 &   12.2 &     \\
4  & 45.83071  &  -15.61531 & 21.93 $\pm$ 0.18 & 21.27 $\pm$ 0.10 & -0.05 $\pm$ 0.20 &   12.2 &     \\
5  & 45.89130  &  -15.62050 & 22.07 $\pm$ 0.18 & 21.63 $\pm$ 0.12 & 0.04  $\pm$ 0.20 &   37.3 &     \\
6  & 45.94355  &  -15.62937 & 21.15 $\pm$ 0.10 & 20.59 $\pm$ 0.06 & 0.20  $\pm$ 0.13 &    &    \\
7  & 45.89273  &  -15.63312 & 21.42 $\pm$ 0.14 & 20.59 $\pm$ 0.06 & 0.52  $\pm$ 0.16 &    &    \\ 
8  & 45.84404  &  -15.64084 & 21.85 $\pm$ 0.17 & 21.11 $\pm$ 0.09 & 0.26  $\pm$ 0.20 &    &    \\ 
9 & 45.83414  &  -15.64766 & 20.70 $\pm$ 0.10 & 20.26 $\pm$ 0.05 & 0.35  $\pm$ 0.15 &   & 29108 \\ 
10 & 45.84662  &  -15.65240 & 22.24 $\pm$ 0.20 & 20.90 $\pm$ 0.09 & 0.54  $\pm$ 0.25 &    &    \\ 
11 & 45.86941  &  -15.66216 & 21.13 $\pm$ 0.11 & 20.08 $\pm$ 0.04 & 0.58  $\pm$ 0.13 &    &    \\ 
12 & 45.84815  &  -15.66981 & 21.60 $\pm$ 0.14 & 21.01 $\pm$ 0.07 & 0.37  $\pm$ 0.15 &    &    \\
13 & 45.87857  &  -15.68500 & 21.29 $\pm$ 0.12 & 19.76 $\pm$ 0.04 & 0.72  $\pm$ 0.15 &   &  9342   \\
\hline
\end{tabular}
\end{minipage}
\end{table*}

\begin{table*}
\begin{minipage}[t]{\textwidth}
\caption{Observed and Derived Properties for HCG 23 Regions}
\label{table6}
\centering
\renewcommand{\footnoterule}{}  % to avoid a line before footnotes
\begin{tabular}{cccccccc}
\hline \hline
ID & R.A.& DEC. & FUV & NUV & FUV-NUV\footnote{Using a fixed apertures of 4\arcsec.} & Age\footnote{Age from FUV-NUV using Thilker et al. (2007).}& Velocity\footnote{Velocity taken from de Carvalho et al. (1997) and Hickson et al. (1992) for object 9 and 2 respectively.}\\
~ &(2000)&(2000) & ~& ~ & ~ &(Myr) &(km s$^{-1}$)\\
\hline
1  & 46.71703  &  -9.56431 & 21.64 $\pm$ 0.11 & 21.02 $\pm$ 0.05 & 0.50  $\pm$ 0.10 &      &  \\
2  & 46.79337  &  -9.57706 & 18.67 $\pm$ 0.03 & 17.86 $\pm$ 0.01 & 0.49  $\pm$ 0.05 &	   &  10150 \\
3  & 46.71273  &  -9.59288 & 19.81 $\pm$ 0.04 & 19.51 $\pm$ 0.02 & 0.41  $\pm$ 0.07 &	   &   \\
4  & 46.71515  &  -9.60920 & 21.55 $\pm$ 0.14 & 21.24 $\pm$ 0.07 & 0.45  $\pm$ 0.13 &	   &   \\
5  & 46.72348  &  -9.62984 & 20.49 $\pm$ 0.07 & 19.31 $\pm$ 0.03 & 0.45  $\pm$ 0.08 &	   &   \\
6  & 46.79189  &  -9.62909 & 21.64 $\pm$ 0.10 & 20.56 $\pm$ 0.03 & 0.82  $\pm$ 0.09 &	   &   \\
7  & 46.81795  &  -9.63047 & 21.30 $\pm$ 0.09 & 21.29 $\pm$ 0.06 & -0.07 $\pm$ 0.09 &	10.0   &    \\
8  & 46.77020  &  -9.67570 & 21.11 $\pm$ 0.09 & 20.87 $\pm$ 0.04 & 0.20  $\pm$ 0.08 &	   &   \\
9  & 46.78895  &  -9.48790 & 19.90 $\pm$ 0.05 & 19.90 $\pm$ 0.02 & 0.00  $\pm$ 0.05 &	 21.5 & 5283 \\
\hline
\end{tabular}
\end{minipage}
\end{table*}

\begin{table*}
\begin{minipage}[t]{\textwidth}
\caption{Observed and Derived Properties for HCG 92 Regions}
\label{table7}
\centering
\renewcommand{\footnoterule}{}  % to avoid a line before footnotes
\begin{tabular}{ccccccc}
\hline \hline
ID & ID\footnote{ID in Mendes de Oliveira et al. (2001,2004).}&R.A.& DEC. & FUV-NUV\footnote{Using a fixed apertures of 4\arcsec.} & Age\footnote{Age from FUV-NUV using Thilker et al. (2007).}& Velocity\footnote{Taken from Mendes de Oliveira et al. (2004).}\\
~ & &(2000)&(2000)  & ~ &(Myr) &(km s$^{-1}$)\\
\hline
1   & ...& 339.11306  & 34.03644  & -0.09 $\pm$ 0.08& 5.1 &     \\
2   & ...& 339.10057  & 34.04839  & 0.73  $\pm$ 0.12&     &   \\
3   & ...& 339.08941  & 33.96354  & 0.53  $\pm$ 0.13&     &      \\
4   & a& 339.06386  & 33.98405  & -0.16 $\pm$ 0.10& 3.4	 &  6600  \\
5   & ...& 339.06090  & 34.02086  & -0.56 $\pm$ 0.03& 0.1	  &   \\
6   & b& 339.05867  & 33.97303  & -0.12 $\pm$ 0.13& 3.9	 &  6651  \\
7   & ...& 339.05790  & 33.96437  & -0.09 $\pm$ 0.11& 5.1	 &     \\
8   & d& 339.04965  & 33.98702  & -0.18 $\pm$ 0.10& 3.2	 &  6616   \\
9   & ...& 339.04965  & 33.95852  & 0.20  $\pm$ 0.07&	 &      \\
10  & ...& 339.04957  & 34.02240  & 0.20  $\pm$ 0.12&	&  \\  
11  & c& 339.04842  & 33.98374  & 0.30  $\pm$ 0.12&	& 6565 \\    
12  & ...& 339.04518  & 34.03890  & 0.37  $\pm$ 0.11&	&  \\   
13  & ...& 339.03988  & 33.93911  & 0.47  $\pm$ 0.11&	& \\   
14  & ...& 339.03426  & 33.98144  & 0.21  $\pm$ 0.13&	& \\   
15  & ...& 339.01353  & 33.99232  & 0.00  $\pm$ 0.13& 21.5	  &    \\   
16  & ...& 339.00492  & 33.89762  & 0.41  $\pm$ 0.07&	&  \\   
17  & ...& 339.00299  & 33.99783  & -0.2  $\pm$ 0.10& 3.1	  &     \\  
18  & 10& 339.00132  & 33.96527  & 0.14  $\pm$ 0.04&	  &    \\   
19  & ...& 339.00029  & 33.92982  & 0.49  $\pm$ 0.07&	  &     \\  
20  & ...& 339.00018  & 33.96802  & 0.19  $\pm$ 0.04&	 &  \\    
21  & 12& 338.99871  & 33.96096  & 0.2   $\pm$ 0.04&	&  \\    
22  & 7& 338.99737  & 33.97522  & 0.16  $\pm$ 0.05&	 &     \\   
23  & ...& 338.99687  & 33.99399  & 0.04  $\pm$ 0.10& 37.3	  &    \\   
24  & 14& 338.99626  & 33.95968  & 0.13  $\pm$ 0.03&	  &    \\   
25  & 13& 338.99283  & 33.98158  & 0.02  $\pm$ 0.03& 30.5	  &    \\   
26  & 5& 338.98729  & 33.96068  & 0.11  $\pm$ 0.05&	   &    \\  
27  & 16& 338.98496  & 33.97171  & 0.05  $\pm$ 0.04& 40.8	    &  \\   
28  & 8& 338.98478  & 33.95249  & -0.07 $\pm$ 0.07& 10.0	    &   \\  
29  & 23& 338.98404  & 33.98615  & 0.25  $\pm$ 0.08&	&  \\    
30  & ...& 338.98385  & 33.94899  & -0.02 $\pm$ 0.08& 19.1	   &  \\    
31  & ...& 338.98203  & 33.96181  & 0.00  $\pm$ 0.03& 21.5	   &   \\   
32  & 18& 338.98033  & 33.95322  & -0.11 $\pm$ 0.05& 4.0	   &   \\   
33  & 22& 338.97658  & 33.95873  & 0.23  $\pm$ 0.08&	  &  \\   
34  & 20& 338.97609  & 33.95446  & -0.02 $\pm$ 0.04& 19.1	   &   \\   
35  & 21& 338.95747  & 33.94368  & 0.55  $\pm$ 0.09&	&   \\   
36  & ...& 338.94114  & 33.91203  & 0.44  $\pm$ 0.11&	 &      \\   
\hline
\end{tabular}
\end{minipage}
\end{table*}

\begin{table*}
\begin{minipage}[t]{\textwidth}
\caption{Observed and Derived Properties for HCG 100 Regions}
\label{table8}
\centering
\renewcommand{\footnoterule}{}  % to avoid a line before footnotes
\begin{tabular}{cccccccc}
\hline \hline
ID & ID\footnote{ID in de Mello et al. (2008a).} &R.A.& DEC. & FUV & NUV & FUV-NUV\footnote{Using a fixed apertures of 4\arcsec.} & Age\footnote{Age from FUV-NUV using Thilker et al. (2007).}\\
~ & &(2000)&(2000) & ~& ~ & ~ &(Myr) \\
\hline
1  & ... &  0.37526 &  13.16465 & 21.90 $\pm$ 0.10 & 21.38 $\pm$ 0.06 & 0.47  $\pm$ 0.10 &	  \\
2  & ... &  0.30563 &  13.17565 & 21.70 $\pm$ 0.08 & 20.29 $\pm$ 0.03 & 0.81  $\pm$ 0.07&	       \\
3  & ... &  0.34740 &  13.16286 & 21.34 $\pm$ 0.07 & 20.31 $\pm$ 0.04 & 0.82  $\pm$ 0.09&	       \\
4\footnote{Recently obtained spectroscopy confirms these regions as part of HCG 100 (Urrutia et al. 2009).}  & 4\footnotesize$^{d}$ &  0.29263 &  13.08581 & 21.39 $\pm$ 0.09 & 20.95 $\pm$ 0.07 & 0.00  $\pm$ 0.10 &   21.49 \\
5  & ... &  0.30592 &  13.14859 & 21.79 $\pm$ 0.11 & 20.70 $\pm$ 0.06 & 0.91  $\pm$ 0.12&	       \\
6  & 6 &  0.30766 &  13.16278 & 20.38 $\pm$ 0.05 & 19.92 $\pm$ 0.03 & 0.23  $\pm$ 0.06&	       \\
7  & 13 &  0.37282 &  13.09862 & 19.65 $\pm$ 0.03 & 19.19 $\pm$ 0.02 & 0.07  $\pm$ 0.05&   48.75    \\
8  & 14 &  0.37734 &  13.10402 & 21.51 $\pm$ 0.08 & 21.12 $\pm$ 0.07 & -0.06 $\pm$ 0.08&   10.86	 \\
... & 3\footnotesize$^{d}$ & 0.27220 & 13.02360 & 21.21 $\pm$ 0.09 & 21.61 $\pm$ 0.08 & -0.40 $\pm$ 0.12 & $<$1 \\
\hline
\end{tabular}
\end{minipage}
\end{table*}

\begin{table*}
\begin{minipage}[t]{\textwidth}
\caption{Observed and Derived Properties for NGC 92 Regions}
\label{table9}
\centering
\renewcommand{\footnoterule}{}  % to avoid a line before footnotes
\begin{tabular}{ccccccccccccccccccc}
\hline \hline
ID & R.A.& DEC. & FUV & NUV & FUV-NUV\footnote{Using a fixed apertures of 4\arcsec.} & Age\footnote{Age from FUV-NUV using Thilker et al. (2007).}& SFR$_{FUV}$\footnote{SFR (M$_{\odot}$ yr$^{-1}$) from Iglesias-P\'aramo et al. (2006) using FUV corrected for internal extinction from Gil de Paz et al. (2007) and Buat et al. (2005). Note that these values have uncertainties of typically a factor of 2, given that we have no measured IR/UV ratios for these objects.} &Velocity\\
~ &(2000)&(2000) & ~& ~ & ~ &(Myr)& (M$_{\odot}$ yr$^{-1}$) & (km s$^{-1}$)\footnote{Estimated from the velocity field}\\
\hline
1 &  5.39251  & -48.63508   & 20.04 $\pm$ 0.03 & 19.67 $\pm$ 0.02 & 0.34 $\pm$ 0.07 &       & 0.008 &  3283  \\
2 &  5.39949  & -48.64607   & 20.55 $\pm$ 0.04 & 20.31 $\pm$ 0.01 & 0.19 $\pm$ 0.06 &       & 0.003 &  3289   \\
3 &  5.41036  & -48.64908   & 18.96 $\pm$ 0.02 & 18.77 $\pm$ 0.01 & 0.06 $\pm$ 0.04 &  44.6 & 0.011 &  3294 \\
4 &  5.32479  & -48.64329   & 21.38 $\pm$ 0.07 & 19.99 $\pm$ 0.03 & 0.43 $\pm$ 0.08 &       &       &      \\
5 &  5.33817  & -48.67298   & 21.75 $\pm$ 0.10 & 21.07 $\pm$ 0.05 & 0.38 $\pm$ 0.12 &       &       &       \\
\hline
\end{tabular}
\end{minipage}
\end{table*}

Considering the relatively small size of our samples and large dispersions in the averages, we have to be 
particularly careful in our analysis. Therefore, instead of comparing average numbers, we performed the Kolmogorov-Smirnov (KS) statistical test comparing 
the color distribution of each group with its corresponding control sample (Table \ref{table10}). 
The most significant result is for HCG 92, where the intragroup FUV--NUV color distribution shows only a 
0.001\% probability of being drawn from the control sample distribution. For the other groups, HCG 2, 
HCG 22, and NGC 92 have distributions similar to 
their control samples, at the 67\%, 47\%, and 96\% level, 
respectively. HCG 7 and HCG 100, on the other hand, 
are similar only at 6\% and 33\% levels respectively 
(note that for HCG 7 regions are on average
redder than the CS regions). If we take the intragroup sample of 
de Mello et al. (2008a) for HCG 100 and its respective control sample, we find only a 6\% probability, which confirms 
that de Mello et al. (2008a) intragroup regions are significantly bluer than regions outside the group.

In order to increase our control sample and improve our statistics, we generated one total control sample by 
adding all control samples (Table \ref{table10}). The average UV color for the total control sample is FUV--NUV=$0.35 \pm 0.25$. This value is slightly lower than the 
peak in the FUV--NUV color distribution given in Gil de Paz et al. (2007) (FUV--NUV$\simeq$0.4) but still consistent with it. This small difference is 
due to our criteria of colors bluer than FUV--NUV=1. When we consider the control sample in the same range of colors as Gil de Paz et al. (2007) (-0.4$\leq$FUV-NUV$\leq$3.2), our average is FUV--NUV$=0.4.$

HCG 92, HCG 22 and NGC 92 have FUV-NUV averages FUV--NUV=$ 0.13\pm0.26$,
FUV--NUV=$ 0.32\pm0.26$, FUV--NUV=$ 0.28\pm0.15$ bluer than the total
control sample (FUV--NUV=$0.35\pm0.25$). When we 
compare each group sample with the total CS, the KS test gives the lowest
probabilities to HCG 92 ($\sim$0\%), HCG 7 (3\%), and HCG 100 (8\% or 32\%
depending if using de Mello et al. (2008a) or the HCG 100 generated in
this work).

Finally, we added all regions of all groups and compared the FUV-NUV distribution with 
that of the total control sample. We find a probability of 0.005\% that CS and group regions are drawn from the same
distribution of colors. However, when we remove from the list all regions belonging to HCG 92 and HCG 100,
the probability increases to 48\%. Therefore, we conclude that this is a clear evidence that these
two groups are significantly different from the other five in this respect. Moreover, as shown in Session 5.3,
these two groups are also unique in their HI properties.

\subsection{Colors, Photometric Ages and SFR}

\begin{figure}[t!]
\centering
\vspace{0.8cm}\includegraphics[scale=0.36]{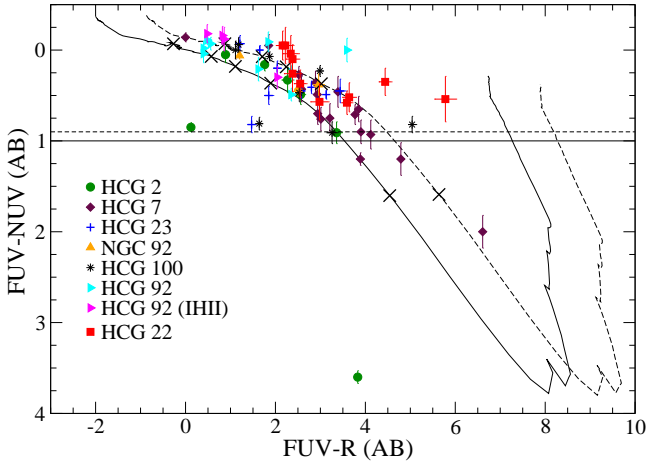}
\caption{\textit{GALEX} FUV-NUV versus FUV-R of the intragroup objects. Models from Thilker et al. (2007) are shown as a solid line (no extinction correction) and dotted line (internal extinction of the Milky Way E(B-V)=0.2). Crosses mark ages 10, 50, 100, 200, and 500 Myr, from the bottom right to the top left. We show only a few objects for HCG 92 due to the small field of view of the optical image compared with the \textit{GALEX} image. Colors bluer than FUV-NUV=0.9 (dashed line) are typical of spiral and late-type galaxies (Gil de Paz et al. 2007). Objects bluer than FUV-NUV=1 (continuous line) were taken into account in our analysis.}
\label{FUV_R_FUV_NUV_2}
\end{figure}

In Figure \ref{FUV_R_FUV_NUV_2} we plot FUV-NUV and FUV-R colors of the intragroup candidates (colors estimated inside a fixed aperture 
of 4$''$ radius) and the stellar population synthesis models given in Thilker et al. (2007). These models are tuned for \textit{GALEX} and optical 
colors, using an instantaneous burst, a solar metallicity and Chabrier IMF. 
We plot models assuming intrinsic colors and reddened colors 
for E(B-V)=0.2 and a Milky Way attenuation law (R$_{V}$=3.1). We note that bluer objects than FUV-NUV$\leq$0.25 are located close to the 
Milky Way attenuation law. On the other hand, objects redder than this value are 
mostly placed between the models with Milky Way extinction law and with no extinction. Gil de Paz et al. (2007), using more than one thousand galaxies, suggested that the attenuation law in the UV is different from a pure Galactic extinction law (Fig. 8b in Gil de Paz et al. 2007). From this fact and from Fig. \ref{FUV_R_FUV_NUV_2}, we conservatively estimated ages of all objects with FUV--NUV$<$0.1 using Thilker et al. (2007) with Milky Way extinction corrections. Tables \ref{table3} to \ref{table9} show the ages for each region. In Fig. \ref{FUV_R_FUV_NUV_2}, we show the intergalactic HII regions (IHII) detected 
in Mendes de Oliveira et al. (2004) where we can see that two of the four spectroscopically confirmed IHII regions can be adjusted by the Milky Way 
extinction law.

For each region with known distance, we estimated the Star Formation Rate (SFR) using the ultraviolet luminosity of each object. In order to correct luminosities by the dust effect, we used our FUV-NUV colors to obtain the total-infrared to FUV ratio (IR/FUV) for each galaxy using equation 11 of Gil de 
Paz et al. (2007), since we do not have infrared data for our galaxies. The A$_{FUV}$ value was then obtained inserting the derived value of IR/FUV, estimated as above, in the equation 2 of Buat et al. (2005). The UV luminosity is a tracer of star formation since most of the UV photons coming from a galaxy are produced by young stars. In this case, the SFR is proportional to the UV luminosity emitted by OB stars, under the assumption that the SFR is approximately constant over this timescale (Iglesias-P\'aramo et al. 2006). The SFR was estimated (Table \ref{table9}) using the equations given by Iglesias-P\'aramo et al. (2006) and the FUV luminosities for each region. Note that the dispersion found in equation 11 of Gil de Paz et al. (2007) is quite high (0.36 dex), and therefore this introduces large uncertainties in the derived values of the UV attenuation and consequently in the determined values of SFRs (listed only for three regions with known redshifts for group NGC 92, see Table 9).

\begin{figure}[t!]
\centering
\vspace{0.8cm}\includegraphics[scale=0.36]{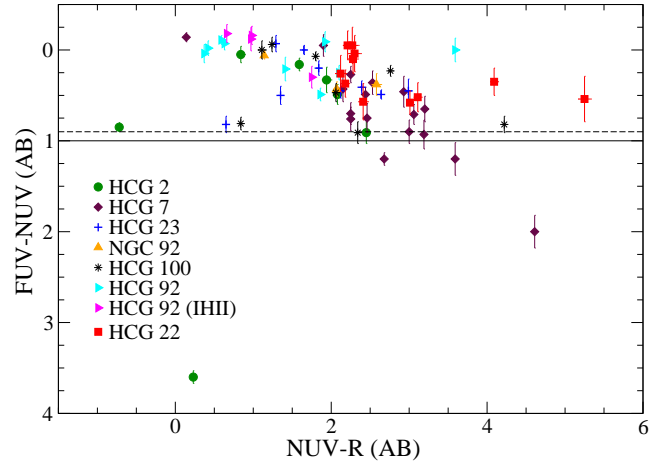}
\caption{FUV-NUV vs NUV-R plot. See caption of Figures \ref{FUV_R_FUV_NUV_2}.}
\label{NUV_R_FUV_NUV}
\end{figure}

Figure \ref{NUV_R_FUV_NUV} shows the NUV--R versus FUV--NUV plot for all regions where we also include objects with colors 
redder than FUV--NUV=1. The reddest object in HCG 2 is outside the region populated by early-type objects (defined by Gil de Paz
et al. 2007) and shows a stellar morphology in our R-band image. Two of the HCG 7 sources also show stellar profile in the r-band image, therefore, 
only one object with color FUV--NUV$\geq$1 shows an extended morphology. In the case of HCG 92, two objects with colors redder than FUV--NUV=1 resemble   
stars (these are objects without R-band information). Another 4 objects in the plot show colors slightly redder than FUV--NUV=0.9 which is the limit
defined by Gil de Paz et al. (2007) as typical of early-type galaxies. However, considering the relatively large error bars they could
also be classified as spiral galaxies. All other intragroup candidates have colors typical of late-type galaxies.

\begin{table}
\begin{minipage}[t]{\columnwidth}
\caption{Statistical values}
\label{table10}
\centering
\renewcommand{\footnoterule}{}  % to avoid a line before footnotes
\begin{tabular}{ccccc}
\hline \hline
~  &\multicolumn{2}{c}{Each Control Sample} &  \multicolumn{2}{c}{Total Control Sample} \\
~  & D\footnote{D is the maximum vertical deviation between two cumulative distribution in the K-S statistics.} & P\footnote{P gives the significance level of the K-S statistic. Small values of P show that the cumulative distribution 
function of CS is significantly different from intragroup regions.} & D & P \\
\hline
HCG 2  & 0.29  & 0.67 & 0.30 & 0.56  \\
HCG 7  & 0.36 & 0.06 & 0.37 &  0.03 \\
HCG 22  & 0.25 & 0.47 & 0.18 & 0.79  \\
HCG 23  & 0.31 & 0.39 & 0.29 & 0.39  \\
HCG 92  & 0.40 & 0.00 & 0.42 & 0.00 \\
HCG 100  & 0.33 & 0.33 & 0.32 & 0.32  \\
HCG 100\footnote{Colors of HCG 100 objects from de Mello et al. (2008a) were estimated using fixed aperture.}  & 0.35 & 0.06 & 0.31 &  0.08 \\
NGC 92  & 0.21 & 0.96 & 0.22 & 0.95 \\
All Regions\footnote{Using HCG 100 detected regions from this work.} &  &  & 0.20 & 0.00 \\
All Regions\footnote{Using HCG 100 detected regions from de Mello et al. (2008a)} &  &  & 0.21 & 0.00 \\
\hline
\end{tabular}
\end{minipage}
\end{table}

\section{Discussion}

\subsection{Kinematics and the evolutionary stage of each group}

We used the scheme devised by Amram et al. (2003) and Plana et al. (2003) to classify galaxies for which we have Fabry-Perot velocity maps 
as interacting or not, based on the analysis of their 2-D velocity fields and the determination of seven interaction indicators. 
The first indicator of interactions is the presence of strong peculiarities in the velocity field, which may display
non-circular motions. A highly peculiar velocity
field, not common for normal galaxies (see e.g. Epinat et al. 2008a, b), is a strong sign of interaction. A perturbed velocity field leads, by consequence, 
to a peculiar rotation curve, except if the asymmetries seen in the velocity field are smoothed by azimuthal averaging. This is the case, for example, for HCG 7a which shows an asymmetric velocity field with respect to the major axis (Fig. \ref{maps_hcg7ad}), completely smoothed on the rotation curve after the azimuthal averaging (Fig. \ref{hcg22c}d). On the other hand, asymmetries with respect to the minor axis of the velocity field are easier to notice by comparing both sides of the rotation curve than by inspecting the velocity field. Therefore, we introduce this specific interaction indicator which refers to asymmetries in the shapes of the rotation curves.
A rotation curve is considered peculiar 
when it is not possible to match the shape of the receding with that of the approaching side, no matter how one chooses
the galaxy kinematic center and systemic velocity. 

The third and fourth indicators refer to the orientation of the PA of
the galaxy. Epinat et al. (2008b) showed that 21\% (33 objects) of field
late-type galaxies from GHASP survey has a misalignment between kinematic and
morphological axes greater than
20$^{\circ}$.  From these 33, if we remove galaxies with low inclinations
and with bad morphological PA determinations, almost all remaining
ones have bars.  This means that for field galaxies (as represented
by the GHASP sample), the misalignment between stellar and kinematic
axes happen quite rarely, except for barred galaxies. The presence of a bar has not been proven observationally to be directly associated with interactions, although several N-body simulation studies of bar formation have suggested that a connection between galaxy-galaxy interaction and bar formation exists and depends on the orbital parameters of the encounter (e.g. Debattista et al.
2006; Gerin et al. 1990). On the other hand, it has been observed that in several non-barred interacting galaxies, there is a clear misalignment between
the gas and stellar PA or a change of PA along the major axes, given that
accreted gas can settle in a different disk (e.g. galaxies in HCG 16,
Mendes de Oliveira et al. 1998, and in HCG 100, Plana et al. 2003). For this
reason we include "Gaseous versus stellar major-axis misalignment" and 
"Changing position
angle along major axis" as interaction indicators in Table 11, although we have to
keep in mind that these indicators may also be related to the existence
of a bar (such cases are flagged in Table 11 with the footnote b).

%The third and fourth indicators refer to the measurement of the position angle of the galaxy. Misalignment between gas and stellar major axis has being widely interpreted as a sign of interactions (e. g. Mendes de Oliveria et al. 1998) as accreted gas settled in a different disk. Nevertheless Epinat et al. (2008b) showed that 21\% of field late-type galaxies from the GHASP survey have a misalignment greater than 20$^{\circ}$. If we remove galaxies with low inclinations and galaxies with a bad morphological determination of the position angle in that 21\% of the GHASP sample, 88\% of the remaining galaxies have bars. In this sense, the presence of a bar is not a direct evidence of interaction, however, bars could be driven by the interaction between galaxies. For this reason we keep the indicator ''Gaseous versus stellar major-axis misalignment''. Interactions can produce changes in the position angle along the major axis of the galaxies, moreover, bars can also mimic this effect. We introduce this indicator as ''changing position angle along amjor axis''.

%A large change in the position angle of the system with radius as well as a misalignment (greater than 20 degrees) between the gaseous and stellar major axes are indicators of interaction. 

Other indicator is the presence of multiple components in the H$\alpha$ profiles. It is used to show evidences for strong interactions or mergers although such profiles could, in general, also be associated
with the presence of an AGN.
Finally, the last 
indicators are related to the existence of tidal tails, if the galaxy has high IR emission or not and the presence of central activity for each galaxy.
A list with these parameters for all galaxies studied here are given in Table \ref{table11}. The first four indicators for members of HCG 23 and the galaxies HCG 22a, NGC 88 and NGC 89 are not included in the 
table because no Fabry-Perot data are available for these objects. The first four indicators for HCG 92 are
also not included because no Fabry Perot velocity maps could be
built given that for this group the ionized gas is outside the galaxies (Mendes de Oliveira et al. 2001, Plana et al. 1999).

By inspecting Table \ref{table11} one can see that three galaxies present disagreement between both sides of the rotation curves, HCG 2a, HCG 2b and
NGC 92. These galaxies have close companions and they may have suffered a recent encounter. This has been modelled by 
Pedrosa et al. (2008) for major mergers and by Kronberger et al. (2006) for fly-bys and minor mergers. In the case of a major merger, Pedrosa et al. 
finds that the significantly disturbed rotation curve occurs in a small time interval within 0.5 Gyr of the occurrence of the pericenter. In the case of NGC 92, this time is consistent with the dynamical time of the tidal tail, $\sim$0.2 Gyr, as we derived from the maximum tidal extent divided by the maximum expected tidal velocity (Hibbard et al. 2005). At the tip of this tail, 
we detected a tidal dwarf galaxy candidate. 
Although the presence of double profiles in the central region  
of this galaxy could be associated with its LINER nucleus, 
double components are also observed along the tidal tail and 
these are not correlated to the H$\alpha$ intensity. 
For these reasons we claim that the double components are most likely 
related to strong interactions.
NGC 92, therefore, may have had a very recent interaction with one of its close neighbours. It is possible that also HCG 2a and HCG 2b had a close encounter.

The kinematics of HCG 7a as a whole, does not show signatures of strong interaction with other members of this group (although its velocity field is peculiar, 
as described in section 4.1.2). No peculiarities 
were found in Mendes de Oliveira et al. (2003) for HCG 7c. Taking into account these kinematic results, members of HCG 7 have not had recent interactions with each other.

In the case of the triplet HCG 22, only one member, HCG 22c, has ionized gas and a symmetric rotation curve was determined. As we note in \S 4.1, HCG 22b has shells, 
which indicates a possible merger between an elliptical and a small disk galaxy (Combes and Charmandaris 1998). The timescale for shell formation is 1-10 Gyr.

Plana et al.(2003) showed that the members of group HCG 100 are in an advanced stage of interaction.

\begin{table*}
\begin{minipage}[t]{\textwidth}
\caption{Interaction Indicators}
\label{table11}
\centering
\renewcommand{\footnoterule}{}  % to avoid a line before footnotes
\begin{tabular}{lccccccc}
\hline \hline
\multicolumn{2}{c}{Interaction Indicator\footnote{REFERENCES.-- The indicator ''High IR luminosity'' was taken from the following sources: Allam et al.(1996), Verdes--Montenegro et al. (1998) and Gallagher et al. (2008). The indicator ''Central Activity'' was taken from: Shimada et al. (2000), de Carvalho et al. (1997), Coziol et al. (1998a), Coziol et al. (2000) and Xu et al. (2003).}} &  \multicolumn{2}{c}{Galaxy} \\ 

        & HCG 2a  & HCG 2b  & HCG 2c &   \\
\hline
Highly disturbed velocity field & + & -- & -- &  \\
Disagreement between both sides of the rotation curve  & + & + & -- &   \\
Gaseous versus stellar major-axis misalignment  & +\footnote{The disagreement between optical and kinematic position angle could be associated with the presence of a bar.} & -- & -- &  \\
Changing position angle along major axis  & + & -- & -- &  \\
Multiple components in the profiles &-- &-- &-- & \\
Tidal Tails  & -- & -- & -- &   \\
High IR luminosity  & -- & + & -- &  \\
Central activity   & + & + & ... &   \\
\hline
        & HCG 7a  & HCG 7b  & HCG 7c & HCG 7d   \\
\hline
Highly disturbed velocity field & + & ... & -- & -- \\
Disagreement between both sides of the rotation curve  & -- & ... & -- & -- \\
Gaseous versus stellar major-axis misalignment  & -- & ... & ... &  +\footnotesize$^{b}$ \\
Changing position angle along major axis  & -- & ... & ... & -- \\
Multiple components in the profiles &-- &... & ... & --\\
Tidal Tails  & -- & -- & -- & --  \\
High IR luminosity  & + & -- & + & -- \\
Central activity   & -- & ... & ... & ... \\
\hline
        & HCG 22a & HCG 22b & HCG 22c &          \\
\hline
Highly disturbed velocity field & ... & ... & -- &  \\
Disagreement between both sides of the rotation curve  & ... & ... & -- &   \\
Gaseous versus stellar major-axis misalignment  & ... & ... & +\footnotesize$^{b}$ &  \\
Changing position angle along major axis  & ... & ... & -- &  \\
Multiple components in the profiles & ...& ...& --& \\
Tidal Tails  & -- & -- & -- &   \\
High IR luminosity  & -- & -- & -- &  \\
Central activity   & + & -- & -- &   \\
\hline
        & HCG 23a & HCG 23b & HCG 23c & HCG 23d  \\
\hline
Tidal Tails  & -- & -- & -- & ...  \\
High IR luminosity  & ... & + & ... & + \\
Central activity   & + & ... & + & --  \\
\hline
        & HCG 92b & HCG 92c & HCG 92d & HCG 92e  \\
\hline
Tidal Tails  & + & + & + & --  \\
High IR luminosity  & + & + & -- & -- \\
Central activity   & -- & + & -- & --  \\
\hline
        & HCG 100a\footnote{From Plana et al. (2003)} & HCG 100b\footnotesize$^{c}$ & HCG 100c\footnotesize$^{c}$ & HCG 100d\footnotesize$^{c}$ \\
\hline
Highly disturbed velocity field & + & + & + & + \\
Disagreement between both sides of the rotation curve  & ... & ... & ... & ...  \\
Gaseous versus stellar major-axis misalignment  & -- & + & + & -- \\
Changing position angle along major axis  & + & + & + & -- \\
Multiple components in the profiles &-- &-- &-- &-- \\
Tidal Tails  & -- & + & -- & --  \\
High IR luminosity  & + & -- & -- & -- \\
Central activity   & -- & -- & -- & --  \\
\hline
        & NGC 92 & NGC 89 & NGC 88 & NGC 87  \\
\hline
Highly disturbed velocity field & -- & ... & ... & ... \\
Disagreement between both sides of the rotation curve  & + & ... & ... & ...  \\
Gaseous versus stellar major-axis misalignment  & -- & ... & ... & ... \\
Changing position angle along major axis  & + & ... & ... & ... \\
Multiple components in the profiles & +&... &... &... \\
Tidal Tails  & + & -- & -- & --  \\
High IR luminosity  & ... & ... & ... & ... \\
Central activity   & + & + & + & --  \\
\hline
\end{tabular}
\end{minipage}
\end{table*}

%\begin{figure}[t!]
%\centering
%\vspace{0.6cm}\includegraphics[scale=0.36]{/home/storres/trabajo/trabajo_chile/all_sample/Num_FUV_NUV/A_FUV/A_FUV_FUV_NUV_buat05.eps} 
%\caption{A$_{FUV}$ extinction versus FUV-NUV. Intragroup regions are identified by different symbols. For each object, A$_{FUV}$ values were estimated as described in the text (see section 4.4). Dotted line, traced line and traced-dotted line show the extinction in the UV expected for the SMC bar, 30 Doradus and MW extinction laws, respectively.}
%\label{A_FUV_FUV_NUV_buat}
%\end{figure}

\subsection{Ultraviolet light as an evolutionary diagnostic} 

Despite the fact that tidal dwarf galaxies have been found in a few interacting systems (e.g. Mendes Oliveira et al.
2001, Duc \& Mirabel 1998), other strongly interacting systems present no significant excess of dwarfs 
(Delgado--Donate et al. 2003). In Table \ref{table12}, we present the confirmed intergalactic HII regions and TDGs taken from the literature. In our study, we found an excess of UV emitting regions in the field of HCG 92 
and HCG 22. HCG 92 shows a bluer FUV--NUV color distribution compared with the control sample which is in 
agreement with the previous studies for this group (Mendes de Oliveira et al. 2001, 2004). HCG 22, however,
shows an intragroup color distribution similar to the control sample distribution, i.e. HCG 22 contains 
an older population of intragroup regions than HCG 92. It is possible that these regions are long-lived TDGs 
formed in an earlier interaction and they were able to survive. Similar results have been reported by Duc et al. (2007) for 
the Virgo cluster. This scenario is compatible with the studies developed by Bournaud \& Duc 2006. They found that 
at least 25\% of the substructures located in tidal tails become long-lived bound objects that typically survive 
more than 2 Gyr. In the same way, Hunsberger et al. (1996) found that at least one-third of the dwarf galaxy 
population in compact groups is formed by galaxy interactions.

Although only HCG 92 and HCG 22 show higher field densities compared with their control samples, HCG 100 shows a blue peak in the FUV--NUV color distribution. This group contains TDGs (de Mello et al. 2008a and Urrutia et al. 2009) 
and has been suggested to be 
in an advanced evolutionary stage by Plana et al. (2003).

NGC 92 has only five intragroup regions, three of them (\#1, \#2 and \#3) are located in the extended tidal tail of NGC 92 and are 
clearly seen in the NUV, FUV and R-band images. We note the presence of a strong UV 
emission close to the end of the tidal tail of NGC 92, similar 
to the UV emission from the Antennae TDG candidate (Hibbard et al. 2005). Our detected region at the end of the tidal tail is only 45 Myr old. 
Bournaud \& Duc (2006) performed simulations in order to relate TDGs and satellite galaxies. They found that the most massive 
objects located at the end of tidal tails are likely to be the TDG progenitors. Therefore, considering its location and age, we classify the object in the 
tidal tail of NGC 92 as a TDG candidate. We also note that this object is 
detected in H$\alpha$ (Fabry Perot image) confirming its group membership. 
As has been noted for the Antennae (Hibbard et al. 2005), we detect a gradient in the colors of 
the regions detected along the tail. The bluest region is the TDG candidate, at the tip of the tail of NGC 92.

HCG 2 and HCG 23 have no excess of UV regions in their field and have color distributions similar to their  
control sample. Although the intragroup regions in HCG 7 and in the CS have different color distribution (KS test, Table \ref{table10}), 
it does not present an excess of intragroup regions in the field 
and moreover the regions in 
HCG 7 are generally redder than the colours of the CS regions. 
We conclude that these groups have not actively formed new regions in the intragroup medium which
might be an indication that they are in an early stage of evolution, i.e. that HCG 2, HCG 7 and 23 are non-evolved groups with respect to the UV properties.

We have also found that HCG 92 and HCG 100 intragroup regions 
have color distributions significantly different from those for other groups.
We used Gil de Paz et al. (2007) relation between the FUV-NUV color of a galaxy and its morphological type to evaluate 
the properties of their intragroup regions. Using equation 5 of Gil de Paz et al. (2007), objects  bluer 
than FUV--NUV=0.23 are typically found to be irregular/compact galaxies (morphological type T$\geq$9.5). Following this criteria, most of 
the objects in HCG 92 and in the blue peak of HCG 100 would have this morphological type, which is expected if some of them are TDG candidates.

What are the effects of dust and the typical ages of these UV-selected intragroup regions?  
The relation between UV and FIR emission can be used to obtain an 
estimate of the dust attenuation (Buat et al. 2005 and references therein). 
In our sample, we found that several star forming regions show blue FUV--NUV colors, such as those for the IHII regions 
in HCG 92 (Mendes de Oliveira et al. 2004). 
If the color of IHII is typical of star-forming regions, the attenuation in those objects is modest, with values lower than A$_{FUV}\sim$1 (see e.g. the extinction laws in Fig. 8b of Gil de Paz et al. 2007). Even though the FUV attenuation is low for this type of objects, they must be 
corrected for this effect. The SFR that we calculated (see Table \ref{table9}) using UV colors corrected for dust attenuation are modest but 
similar to the values found by Duc \& Mirabel (1998) for some TDG candidates (see Table \ref{table12}).

We estimated ages using UV colors and Milky Way extinction law (Thilker et. al 2007) only for the bluest objects (FUV--NUV$<$0.1) in our sample (Tables \ref{table3}-\ref{table9}). We compared the age values we estimated with the ones independently estimated from spectroscopy by Mendes de Oliveira et al. (2004, regions \#4, \#6 and \#8 in our Table \ref{table7}) and found an excellent agreement in three of the four regions, showing ages $<$ 10 Myr. The other IHII region (\#11 in Table \ref{table7}) has FUV--NUV$\sim$0.3 which is redder than
FUV--NUV$<$0.1, limit used to estimate ages using a Milky Way extinction law. 

\begin{table*}
\begin{minipage}[t]{\textwidth}
\caption{Other Intergalactic HII regions and TDGs listed in the literature}
\label{table12}
\centering
\renewcommand{\footnoterule}{}  % to avoid a line before footnotes
\begin{tabular}{ccccccccc}
\hline \hline
System & ID & M$_{B}$ & Velocity & 12+log(O/H) & Mass & SFR & IHII-TDG & Reference\footnote{REFERENCES.- 1: Mendes de Oliveira et al. (2004).
2: Mendes de Oliveira et al. (2001), 3: Ryan-Weber et al. (2004), 4: Duc \&
Mirabel (1998). The mass for the candidates in NGC 5291 correspond to the HI mass.}\\
 &  &  & (km s$^{-1}$) &  & (M$_{\odot}$) & M$_{\odot}$ yr$^{-1}$ &  & \\
\hline
HCG 92  & a & -11.9 & 6600 & 8.48 & 2.0x10$^{4}$& &IHII &1\\
HCG 92  & b & -12.1 & 6651 & 8.65 & 1.4x10$^{4}$& &IHII &1\\
HCG 92  & c & -12.5 & 6565 & 8.64 & 4.4x10$^{4}$& &IHII &1\\
HCG 92  & d & -12.3 & 6616 & 8.53 & 3.6x10$^{4}$& &IHII &1\\
HCG 92  & 2 & -14.0 & & & 38x10$^{8}$ & 4.26 & TDG & 2\\
HCG 92  & 3 & -14.7 & & &  &... &TDG & 2\\
HCG 92  & 4 & -13.4 & & &  & ...&TDG & 2\\
HCG 92  & 5 & -15.3 & & &  & ...&TDG & 2\\
HCG 92  & 6 & -15.0 & & & 20x10$^{8}$ & ...&TDG & 2\\
HCG 92  & 8 & -14.2 & & & 138x10$^{8}$ & 0.38&TDG & 2\\
HCG 92  & 20 & -12.6 & & & 6.7x10$^{8}$ & 0.45&TDG & 2\\
HCG 92  & 21 & -14.4 & & & 24x10$^{8}$ & 0.73 &TDG & 2\\
HCG 92  & 22 & -14.2 & & & 4.3x10$^{8}$ & 0.63&TDG & 2\\
HCG 92  & 23 & -13.0 & & & 2.4x10$^{8}$ & 0.20&TDG& 2\\
HCG 16 	       & 1 &  & 3634 & & & &IHII& 3\\
ESO 154-G023   & 1 &  & ... & & & &IHII& 3\\
NGC 1314       & 1 &  & ... & & & &IHII& 3\\
NGC 1533       & 1 &  & 846 & & & &IHII& 3\\
NGC 1533       & 2 &  & 831 & &  & &IHII& 3\\
NGC 1533       & 3 &  & ... & &  & &IHII& 3\\
NGC 1533       & 4 &  & ... & &  & &IHII& 3\\
NGC 1533       & 5 &  & 901 & &  & &IHII& 3\\
IC 5052        & 1 &  & ... & &  & &IHII& 3\\
IC 5052        & 2 &  & ... & &  & &IHII& 3\\
ESO 238-G005   & 1 &  & ... & &  & &IHII& 3\\
ESO 238-G005   & 2 &  &...  & &  & &IHII& 3\\
ESO 149-G003   & 1 &  & 949 & &  & &IHII& 3\\
NGC 5291  & a & -15.54 & 4642 & 8.46 &  0.91x10$^{9}$ & 0.031&TDG& 4\\
NGC 5291  &b  & -14.13 & 4536 & 8.39 &  0.13x10$^{9}$ & 0.004&TDG& 4\\
NGC 5291  & c & -12.13 & 4617 & 8.57 &  ... & 0.002&TDG& 4\\
NGC 5291  & d & -13.83 & 4571 & 8.41 &  0.54x10$^{9}$ & 0.001&TDG& 4\\
NGC 5291  & e & -14.57 & 4415 & 8.57 &  0.43x10$^{9}$ & 0.002&TDG& 4\\
NGC 5291  & f & -13.83 & 4035 & 8.63 &  ... & 0.002&TDG& 4\\
NGC 5291  & g & -14.16 & 4162 & 8.47 &  ... & 0.012&TDG& 4\\
NGC 5291  & h & -13.98 & 4098 & 8.46 &  ... & 0.005&TDG& 4\\
NGC 5291  & i & -16.38 & 3996 & 8.38 &  2.48x10$^{9}$& 0.063&TDG& 4\\
NGC 5291  & j & -13.74 & 4075 & 8.48 &  ... & 0.007&TDG& 4\\
NGC 5291  & k & .13.40 & 4041 & 8.45 &  ... & 0.002&TDG& 4\\
\hline
\end{tabular}
\end{minipage}
\end{table*}

\subsection{The H{\sc i} gas}

It is well known that interactions can cause severe disruption in the HI distribution and produce HI tidal
tails which might host TDGs (Neff et al. 2005). Moreover, there are cases where HI is also found within clouds in 
the intergalactic medium (e.g. Williams et al. 2001) which were removed from galaxies' disks by 
tidal forces. Therefore, HI morphology and content may be a trace of evolutionary stage of interacting systems. In order to
quantify this effect, Verdes-Montenegro et al. (2001) analyzed the HI content of 72 compact groups of galaxies. 
They defined the HI deficiency in a group as $Def_{HI}=log[M(HI)_{pred}]-log[M(HI)_{obs}]$, where the predicted mass for the group is 
the sum of the expected mass for each galaxy member. They proposed the following  
evolutionary sequence.  $''$Phase 1$''$, when the HI distribution in groups  is relatively not perturbed, and almost all 
the HI gas is located in the disk of the galaxies with the remaining gas 
found in incipient tidal tails; $''$Phase 2$''$, when 30 to 60\% of the total HI mass forms tidal features; $''$Phase 3a$''$, 
when the HI is located in tidal tails or when there is no HI; and $''$Phase 3b$''$, when the HI gas is located in a large 
cloud encompassing all galaxies. 

We obtained HI information for each group in our sample from the literature and evaluated their properties taking
into account Verdes-Montenegro et al. (2001) scenario as described below.

HCG 2 shows no HI deficiency, $Def_{HI}=0$, and the only interaction 
signature is the incipient HI tails arising from two of its members. For HCG 7, on the other hand, $\sim$80\% of the expected HI is missing (Verdes-Montenegro et al 2001).

For HCG 22, 72\% of the expected HI is missing (Verdes-Montenegro et al 2001) and only HCG 22c shows HI emission. 
Price et al. (2000) noted that the strongest concentration of HI in the distribution is on the side towards 
the elliptical galaxy, although there are no obvious tidal tail features.

HCG 23 is classified as a group in $''$Phase 1$''$ by Verdes-Montenegro et al. (2001). The observed and expected HI content
is the same. Most of the HI in this group is associated with individual galaxies, as noted by Williams \& van 
Gorkom (1995). In the same study, they noted five additional members that lie outside the compact 
group. Velocity maps for galaxies a, b and d show an asymmetry in the motion of the gas with respect to the optical center. 
They suggest that this group is part of a much larger system.

The well studied HCG 92 shows two galaxies with the highest HI deficiency in the Verdes--Montenegro's sample: 
HCG 92b and HCG 92c. This group is classified as being in  $''$Phase 3a$''$ and shows most of the HI ($\sim$33\% of the expected HI) in 
intergalactic clouds and tidal tails.

de Mello et al. (2008a) and Urrutia et al. (2009) 
studied HCG 100 and detected TDG candidates in the extended tidal tail visible only in HI and not in the optical. 
In this group 
only 32\% of the expected HI is detected (Verdes--Montenegro et al. 2001). 

Pompei et al. (2007) classified NGC 92 as a group in $''$Phase 3b$''$. 
Only 32\% of the expected HI was observed in this group ($Def_{HI}=0.51$). 
An extended HI tail follows the optical/UV tail.

From the HI properties for each group, we conclude that galaxies in HCG 2 and HCG 23 does not show clear signatures of interaction. 
On the other hand, HCG 92 and HCG 100 show large tidal tails of neutral gas, far away from the main body of the galaxies.

\subsection{Evolutionary Stage of each Group}

In Table \ref{table11} we list some of the main interaction indicators for each galaxy and in Table \ref{table13} we summarize the signatures 
of evolution for each group using UV, HI and kinematic information. For instance, groups with an excess of UV regions, disturbed velocity fields and HI outside galaxies were considered as highly evolved. 

Although HCG 2 do not show an excess of intragroup regions neither a deficiency in the HI content, the velocity field of HCG 2a shows several signatures of interactions, which may be produced by an interaction with its companion HCG 2b. 

HCG 23 shows no UV excess and no HI gas in the intragroup medium suggesting that this group is in an early-stage of evolution.

On the other hand, HCG 92 presents an evolved evolutionary stage in the UV 
analysis, with several star-forming regions in the intragroup medium confirmed spectroscopically  
(Mendes de Oliveira et al. 2004). It also contains HI in the intragroup medium.

HCG 100 also shows an advanced evolutionary stage. 
Plana  et al. (2003) found strong signatures of interaction in their galaxy 
members, de Mello et al. (2008a) found some TDG candidates in the HI tail of HCG 100. 

Although only $\sim$20\% of the expected HI is observed in HCG 7, there is no excess in the field density of this group 
neither a significantly blue distribution in its color FUV--NUV distribution neither a disturbed kinematics of their galaxy members. This result suggests that the missing HI has not been used to form 
stars in the intragroup medium. On the other hand, Verdes--Montenegro et al. (2001) noted that it is possible to explain the 
HI deficiency in compact groups by tidal removal of the outer HI disk in the galaxies, which can explain the deficiency of 
HI in HCG 7 without subsequent star formation.

The evolutionary stage of HCG 22 is not as clear as for the other targets. If the excess of regions in HCG 22 is the result 
of previous galaxy interactions or merger events, a fraction of the HI gas might have been converted into
stars. This scenario is supported by the HI observations, since only 28\% of the expected gas is observed in this group. 
However, this group contains an elliptical galaxy, HCG 22a, which contains no signatures of mergers (da Rocha et al. 2002, 
de Souza et al. 2004). Moreover, HCG 22b shows shells structures, associated with mergers events. Therefore, if the intragroup regions belong to HCG 22, we suggest that they are formed from galaxy interaction and/or a merger event.

In NGC 92 we did not find any excess of UV regions in their field. However, galaxy NGC 92 shows an extended tidal tail and both sides of the rotation curve do not match. A TDG candidate is clearly detected in our multiwavelength
study, although no rotation has been measured, from our Fabry Perot
data". 
HI gas has been used up in the process of forming stars, which might be one of the reasons why 69\% of the HI is missing in this group.

Finally, we compare our results with another tracer of evolution in groups: X-ray observations. In the X-ray Atlas of Groups of Galaxies, Mulchaey et al. (2003) showed that HCG 92 shows a clear difusse emision (i.e a hot intragroup medium). On the other hand, HCG 2, HCG 22 and HCG 23 did not present a hot intragroup medium (Mulchaey et al. 2003). Interestigly, the spiral rich group NGC 92 has a hot intragroup medium, as was reported by Trinchieri et al. (2008). These results of X-ray emission are in agreement with our clasification for the evolutionary stage of compact groups.

\begin{table}
\begin{minipage}[t]{\columnwidth}
\caption{Evolutionary Stages for the Compact Groups}\footnotetext{For each interaction tracer, symbol + represents a more evolved group. Symbol - means a less evolved group. The + sign in the FUV-NUV column of NGC 92 refers to the blue color of the candiate TDG.}
\label{table13}
\centering
\renewcommand{\footnoterule}{}  % to avoid a line before footnotes
\begin{tabular}{cccc}
\hline \hline
Group & Velocity Field & HI & FUV-NUV\\ 
\hline
HCG 2   & +/-- & -- & -- \\
HCG 7   & -- & + & -- \\
HCG 22  & -- & + & + \\
HCG 23  & ... & -- & -- \\
HCG 92  & + & + & + \\
HCG 100 & + & + & + \\
NGC 92  & + & + & + \\
\hline
\end{tabular}
\end{minipage}
\end{table}

\section{Conclusions}

We present new ultraviolet and new Fabry-Perot data for a sample of 5 compact groups of galaxies. 
For each galaxy, we present UV images, H$\alpha$ velocity field, H$\alpha$ monochromatic image, velocity dispersion map and rotation curve.

We classified the groups according to their evolutionary stages and in addition we searched for young 
star-forming region candidates in galaxies with tidal tails and in the intragroup medium of the compact groups. Dozens of regions with ages $<$ few Myrs were found. 
We concluded that HCG 7 and HCG 23 are non-evolved groups. HCG 2 and HCG 22 show limited signatures of interaction. For instance, HCG 2a has a non-symmetric rotation curve and
HCG 22 has an excess in the field density of star-forming regions and shells structures. We therefore classified both groups as being in a mildly 
interacting stage of evolution. 
We conclude that three groups, HCG 92, HCG 100 and NGC 92, are in advanced stages of interaction, contain extended HI tails and  
harbor young star-forming regions and TDGs. We are in the process of obtaining spectroscopy of a sub-sample of the intragroup objects in order to confirm their group membership. Preliminary results confirm that at least two of the UV-bright sources of HCG 100 (as listed in Table 8) are in the same redshift as the group members (Urrutia et al. 2009).

\begin{acknowledgements}
We would like to thank the referee, Dr. Sonia Temporin, for very useful comments and suggestions which were very important to improve this paper. S. T--F. acknowledges the financial support of FAPESP through the Doctoral position, under contract 2007/07973-3. %S. T--F. would also like to thank the NASA's Goodard Space Flight Center and The Catholic University of America for their hospitality during the summer of 2008. 
C. M. d. O. acknowledges support from the Brazilian agencies FAPESP (projeto tem\'atico 2006/56213-9), 
 CNPq and CAPES. D.FdM acknowledges support from \textit{GALEX} grant NNG06GG45G. HP acknowledges the financial support of CAPES through the  
Pos-Doctoral position, under contract 3656/08-0.
 \textit{GALEX} is a NASA Small Explorer, launched in 2003 April. We gratefully acknowledge NASA's support for construction, 
 operation, and science analysis for the \textit{GALEX} mission, developed in cooperation with the Centre National d' Etudes Spatiales of France and the Korean Ministry of Science and Technology.
This research has made use of the NASA/IPAC Extragalactic Database (NED) which is operated by the Jet Propulsion Laboratory, California Institute of Technology, 
under contract with the National Aeronautics and Space Administration. We also acknowledge the usage of the HyperLeda database (http://leda.univ-lyon1.fr).
\end{acknowledgements}

\end{document}